\documentclass[12pt]{article}\usepackage{amsmath,amssymb,multirow,jheppub2,graphicx}
\usepackage{centernot}
\usepackage{mciteplus}
\usepackage{bbm}
\usepackage{color}
\allowdisplaybreaks[1]
\usepackage{tikz}
 \tikzset{node distance=2cm, auto}

\def\tr{\text{tr}}

\def\tfrac#1#2{{\textstyle{\frac{#1}{#2}}}}

\def\hat{\widehat}
\def\bar{\overline}

\def\be{\begin{equation}}
\def\ee{\end{equation}}
\def\Z{{\mathbb Z}}
\def\R{{\mathbb R}}
\def\coeff#1#2{{\textstyle {\frac {#1}{#2}}}}

\def\half{\coeff 12}

\usepackage{verbatim}   
\usepackage[all]{xy}
\usepackage{bm}  
\def\Dslash{{\rlap{\raise 1pt \hbox{$\>/$}}D}}
\def\Pslash{{\rlap{\raise  1pt \hbox{$\>/$}}\,\partial}}

\title{Critical behavior of gauge theories and Coulomb gases in three and four dimensions} 
 
\author[a,c]{Aleksey Cherman\,} 
\author[b,c]{\!\!,  Mithat \"Unsal\,}
\emailAdd{aleksey.cherman.physics@gmail.com}
\emailAdd{unsal.mithat@gmail.com}
\affiliation[a]{Institute for Nuclear Theory, University of Washington, Seattle, WA USA}
\affiliation[b]{Department of Physics, North Carolina State University,  Raleigh, NC 27695, USA}
\affiliation[c]{Kavli Institute for Theoretical Physics University of California, Santa Barbara, CA 93106}
\abstract{Gauge theories with matter often have critical regions in their parameter space where gapless degrees of freedom emerge.  Using controlled semiclassical calculations, we explore such critical regions in $SU(N)$ gauge theories with a topological $\theta$ term and $N_F$ fundamental fermions in four dimensions, as well as related  field theories in three dimensions.  In  four-dimensional theories, we find that for all $N_F \ge 1$ the critical behavior always occurs at a point in parameter space. For $N_F>1$ this is consistent with the standard QCD expectations,  while for $N_F=1$ our results are consistent with recent observations concerning 't Hooft anomalies. We also show how the $N$-branched structure of observables transmutes into the $N_F$-branched structure seen in chiral Lagrangians as the mass parameter is dialed. As a side benefit, our analysis of these 4D theories implies  the unexpected result that 3D Coulomb gases can have gapless critical points.  We also consider QCD-like parity-invariant theories in three dimensions, and find that their critical behavior is quite different.  In particular, we show that their gapless region is an interval in parameter space, rather than a point.  Our results have non-trivial implications for the infrared behavior of three-dimensional compact QED.   }
  
 \preprint{INT-PUB-17-050 } 
  
\begin{document}
\maketitle


\section{Introduction and results}

This work derives new  non-perturbative results in 4D and 3D $SU(N)$ QCD with $N_F \le N$ fundamental Dirac fermions using  using techniques such as calculable compactifications, adiabatic continuity  and mixed discrete  't Hooft anomalies.  Our main results  are:
\begin{itemize}
\item[1)] It was recently argued that with some natural assumptions, 4D QCD with $N_F=1$ fundamental Dirac fermions has a critical point in the complex mass plane\cite{Gaiotto:2017tne}.  We show that this is indeed the case in a calculable regime of the theory.  This result follows from the existence of a gapless critical point in a 3d Coulomb gas.  This is  an interesting finding in its own right, because historically, it had been believed that 3d  Coulomb gases are always gapped \cite{Polyakov:1976fu, 0022-3719-10-19-011,Read:1990zza, Fradkin:1991nr}.

\item[2)]   Strong coupling extrapolation of weak coupling  chiral symmetry breaking in multi-flavor QCD\cite{Cherman:2016hcd} 
via a  mixed  't Hooft anomaly involving the recently introduced color-flavor center-symmetry\cite{Cherman:2017tey} and chiral symmetry.

\item[3)] 
The transmutation of  $\frac{\theta + 2 \pi k}{N_F} $ $\theta$-angle dependence in the chiral Lagrangian to the  $\frac{\theta + 2 \pi k}{N} $ dependence in pure Yang-Mills as the mass of the quarks are sent to infinity.  

\item[4)] 
The existence of a  {\it critical interval} (rather than a point) in 3d QCD.  
\end{itemize}

We will now summarize our results in more detail, as well as place them in context relative to the various recent developments that have made them possible. 

For all $N_F$ in the range we have studied, we find that in 4D QCD there is a single gapless point in the parameter space of the theory at zero temperature and density.   For $N_F>1$, this is of course entirely expected, because this critical point is at the massless quark point, $m_q = 0$, and is associated with spontaneous chiral symmetry breaking.  For $N_F=1$, the existence of a critical point may sound somewhat surprising, because there is no chiral symmetry breaking reason to expect a gapless mode.  The lightest pseudoscalar meson, the $\eta'$, has historically been expected to have mass $\sim \Lambda_{QCD}$ for any quark mass $m_q$.  However, if one turns on a $\theta$ term, the quark mass is naturally intepreted as a complex parameter, $\arg m_q = \theta$.   Reference~\cite{Gaiotto:2017tne} recently pointed out there is very likely to be a point $m_q \in \mathbb{R}^{-}$ (amounting to $\theta = \pi$) where the $\eta'$ mass must vanish.  This conclusions follows from some very plausible assumptions about the CP-symmetry-breaking behavior of YM theory as a function of $\theta$ as well as anomaly constraints\cite{Gaiotto:2017yup}.
For $m_q < m_{q}^{*}$, CP symmetry is expected to break spontaneously, while it is preserved for $m_q >m_{q}^{*}$.  One can interpret $m_q^{*}$ as a second-order critical point.

We would like to get some microscopic insight into this result, and related results for $N_F>1$.  Our tool will be the idea of adiabatic continuity.  Over the last several years, a large number of results suggest that it is possible to continuously (that is, adiabatically) connect QCD-like theories on $\mathbb{R}^4$ to a class of theories which have a weakly-coupled regime on $\R^3 \times S^1$\cite{Unsal:2007vu,Unsal:2007jx,Unsal:2008ch,Shifman:2008ja,Shifman:2009tp,Unsal:2010qh,Shifman:2008cx,
Cossu:2009sq,Myers:2009df,Simic:2010sv,Vairinhos:2011gv,
Thomas:2011ee,Anber:2011gn,
    Poppitz:2012sw,
    Poppitz:2012nz,Unsal:2012zj,Argyres:2012ka,Argyres:2012vv,
    Anber:2013doa,Cossu:2013ora,Bhoonah:2014gpa,
    Anber:2014lba,Bergner:2014dua,Li:2014lza,
    Anber:2015kea,Anber:2015wha,Misumi:2014raa,
    Cherman:2016hcd,Aitken:2017ayq,Anber:2017rch}.  This weakly-coupled regime appears when the  $S^1$ circumference $L$ is sufficiently small, and (approximate) center symmetry is stable. 
     In the weakly-coupled regime we can use semiclassical methods to determine the behavior of the theory.  We use this as a tool to explicitly see how the $\eta'$ mass manages to vanish at $m_q = m_q^{*}$, in a calculable setting. 
     Our semi-classical analysis on small $S^1$ with appropriate stabilized symmetries (either center or color-flavor-center) is based on a systematic microscopic derivation of chiral symmetry breaking\cite{Cherman:2016hcd}. Our results are consistent with expectations from experimental observations for large $S^1$ and from continuous and discrete 't Hooft anomaly considerations.   

A interesting aspect of this analysis  is that it yields a somewhat unexpected result concerning  Coulomb gases in three dimensions.  Historically it has been believed that such plasmas always have a finite correlation length\cite{Polyakov:1976fu, 0022-3719-10-19-011,Read:1990zza, Fradkin:1991nr}. But we show that our $N_F=1$ QCD results on $\R^3 \times S^1 $ imply the existence of a  gapless point in 3d Coulomb gases in the presence of special complex fugacities for the monopole events. In general, it is known that complex fugacities  induced by Berry phases or topological $\theta$ angle can decrease the mass gap exponentially, from $e^{-S_0}$ down to 
$e^{-2S_0}$ , $e^{-3S_0}$, or $e^{-4S_0}$, see, e.g.  studies of VBS phase of quantum anti-ferromagnets \cite{Read:1990zza, Fradkin:1991nr}   and  deformed Yang-Mills at $\theta = \pi$ \cite{Unsal:2012zj}. 
Our work is the first demonstration that the mass gap in magnetic Coulomb gas  can actually \emph{vanish} at a point in parameter space.

 For $2 \leq  N_F \leq N$,  in the calculable weak coupling regime on $\R^3 \times S^1$ we find that  the critical  point is at $m_q^*=0$,  consistent with QCD expectations on $\R^4$. In this regime, we leverage insights from our  recent work with Thomas Sch\"afer \cite{Cherman:2016hcd}, where the chiral Lagrangian was derived rigorously in the small $\R^3 \times S^1$ setting by using duality and monopole operators.  We find that the theory has a two-fold vacuum degeneracy for $m_q e^{i k\pi/N_F}, k=1, \ldots, N_F$, and a unique vacuum otherwise so long as $m_q \neq 0$.  At $m_q = 0$, the theory has a critical point with gapless excitations.  This behavior is associated with spontaneous continuous chiral symmetry breaking.
  In our set-up,  we can dial the quark mass from small  to large values while keeping all approximations under control.     
  We use this to show how the transmutation of the $\theta$ angle dependence  of observables
   \begin{align} 
   \underbrace{  \frac{\theta + 2 \pi k}{N_F}   }_{\textrm{small $m_q$}}
    \; \longleftrightarrow\;
    \underbrace{ \frac{\theta + 2 \pi k}{N}}_{\textrm{large $m_q$}}
     \end{align}
takes place.  This  $\theta$-dependence transmutation  occurs  as  $m_q$ is taken  from small values, where the softly broken chiral Lagrangian is a good description of the physics,  to  large  values, where the physics becomes describable in terms of pure Yang-Mills theory.  
     
  For the $N_F=2$ theory, the analysis of  \cite{Gaiotto:2017tne} shows that the potential of the leading order chiral Lagrangian vanishes at $\theta=\pi$, and it has not been shown whether the $SU(2)$ flavor symmetry is broken, or CP symmetry is broken. In the calculable regime on $\R^3 \times S^1$, the potential of the monopole-induced chiral Lagrangian vanishes at $\theta = \pi$ when $N_F = 2$. This corresponds to the expected vanishing of the leading-order chiral Lagrangian.  But there are also magnetic-bion induced terms, corresponding to higher-order terms in the chiral expansion.  At small $S^1$, the coefficients of these higher-order terms are calculable, and hence the symmetry breaking pattern can be fixed.  In particular, we find that  CP is broken at $\theta=\pi$.

 For  $2 \leq  N_F \leq N$ and $m_q =0$,  we show that there is a mixed 't Hooft anomaly between color-flavor center (CFC)  symmetry  $\Z_{\rm gcd(N_F, N)}$ \cite{Cherman:2017tey} (assuming  $\gcd(N_F, N) > 1$)   and a $\Z_{2N_F}$ discrete chiral symmetry.   In this set-up,  where $\Z_{\rm gcd(N_F, N)}$  is an exact symmetry on  $\R^3 \times S^1$, the global symmetry $G \sim SU(N_F)_L \times SU(N_F)_R \times U(1)_Q$ is explicitly reduced to its  maximal Abelian subgroup
\begin{align}
 G_{\rm max-ab} = \frac{U(1)^{N_F-1}_V \times U(1)_A^{N_F-1} \times U(1)_Q}{\Z_{N_F}\times \Z_N} \times S_{N_F}  \times \Z_{2N_F}\,.
\label{max-ab}
\end{align} 
by the quark boundary conditions on $S^1$.   Suppose that CFC symmetry is preserved\footnote{CFC symmetry can be stabilized for small $S^1$ using e.g. certain double-trace deformations\cite{Unsal:2008ch}. } when $S^1$ is small, as well as when $S^1$ is large\cite{Iritani:2015ara}. There are no order parameters which transform under $\Z_{2N_F}$, but are neutral under the continuous chiral symmetry $U(1)_A^{N_F-1}$.  Therefore, if the CFC symmetry is unbroken for any $S^1$ size, then mixed discrete 't Hooft anomaly matching actually implies that the \emph{continuous} symmetry $U(1)_A^{N_F-1}$ must be spontaneously broken. This is beautifully 
  consistent with our weak-coupling derivation of chiral symmetry breaking in a regime on $\mathbb{R}^3 \times S^1$ with unbroken CFC symmetry\cite{Cherman:2016hcd}.
The existence of the mixed   't Hooft anomaly imply that our results extrapolate to the strong coupling $\mathbb{R}^4$ regime, without phase transitions for any $S^1$ size, provided that CFC symmetry is unbroken.

Finally,  in order to emphasis the distinction between the above locally four-dimensional analysis, where the dynamics only becomes three-dimensional at long distances, and the dynamics of genuinely three-dimensional theories, we also analyze a three-dimensional $SU(2)$ gauge theory with $N_F = 2$ Dirac fermions.   (This is roughly the dimensional reduction of the 4d $N_F = 1$ Dirac flavor  theory.)  We consider the behavior of this theory as a function of a parity-invariant mass term, and find that its behavior is very different from its four-dimensional cousins.  In particular, in three dimensions, instead of finding a critical point in mass-parameter space, we find  a \emph{critical interval}.   In our three-dimensional set-up, in contrast to an analogous 4D theory, the gapless mode seen in the critical interval is protected by a spontaneously broken continuous shift symmetry, and it is a Nambu-Goldstone boson.  Our results are consistent with recent conjectures discussed in \cite{Komargodski:2017keh}.

\section{$N_F=1$ QCD}
In this section we discuss the critical behavior of one-flavor QCD as a function of the complex quark mass $m_q$ and the number of colors  $N$ by using semi-classical methods and adibatic continuity on $\R^3 \times S^1$.  The Lagrangian is 
\begin{equation}
S= \int_{R_3 \times S_1} \frac{1}{g^2} \left[ \frac{1}{2}\,  \tr  \,
F_{MN}^2  +    i  \frac{ \theta}{16 \pi^2}  \tr  \, F_{MN} \tilde F_{MN} 
+  i  \bar  \Psi  \Dslash 
 \Psi  + m_q  \bar \Psi  \Psi 
    \right]
\label{eq:cont}
\end{equation}
where $\Psi$  is the four-dimensional Dirac spinor in the fundamental 
representation of $SU(N)$, and can be written as
$
\Psi= \left(\begin{array}{l} 
                   \psi_L \cr 
                   \bar \psi_R
                    \end{array}
 \right) $
where $\psi_L,\,\,  \psi_R$ are two-component (complex) Weyl
spinors.
 
This physics of this theory depends on the combination $m = m_q e^{i \theta}$ instead of depending on $m_q$ and $\theta$ separately.   
In fact, one can remove the topological term by a chiral rotation, and write the mass term as 
\begin{align}
m_q e^{i \theta}  \psi_L \psi_R + c.c = m_q \cos \theta  \bar \Psi  \Psi  +  m_q \sin \theta i   \bar \Psi  \gamma_5 \Psi 
\end{align} 
where, on the right hand side  the first mass term is the parity-even Dirac mass term, while the second term is the parity-odd Dirac mass term. 
The  latter representation makes is clear that   $m  \in  \R =  \R^+ \cup   \R^- $ axis is special, in that it corresponds to the presence of CP symmetry, and amounts to setting $\theta=0$ or $\theta = \pi$.  

This theory has a $U(1)_Q/\Z_{N}$  vector-like symmetry, where the $\Z_{N}$ quotient is associated with color gauge transformations living in the center of $SU(N)$.
For $m_q=0$,  the chiral anomaly explicitly breaks the classical $U(1)_A$ symmetry $\Psi \to e^{i \phi \gamma_5} \Psi$ to $\mathbb{Z}_2 = (-1)^{F}$, which is actually a symmetry of the theory with a bare mass. 
The  instanton amplitude is 
 \begin{align}
 I_{4d}  \sim e^{-S_I} \psi_L \psi_R, \qquad S_{I} = \frac{8\pi^2}{g^2}
 \label{ins-mass}
\end{align} 
Unlike $N_F \geq 2$ theories, in which one has an anomaly-free chiral symmetry at $m_q=0$, in the $N_F=1$ theory there is no extra symmetry at  $m_q=0$, mainly due to instanton induced soft mass term \eqref{ins-mass}.   Thus, as a result of non-perturbative effects, the quark mass term receives additive renormalization as opposed to multiplicative renormalization.  In the semi-classical domain, we will use a formalism where we have full control over the additive non-perturbative corrections to the mass term.    

Below, we study the dynamics of this theory on  small $L$ on $\R^3 \times S^1$,  covering $N=2$ first,  and then discuss generic values of $N$ with some comments on large $N$ limit.  All of our analysis assumes that, as mentioned in the introduction, the approximate center symmetry has been stabilized by double-trace deformations.

\subsection{$N=2$ colors}
First, let us consider two-color QCD with quarks of a single flavor.  
In both pure Yang-Mills theory and one-flavor QCD at large $L$, the distribution of the eigenvalues of the color holonomy becomes approximately center-symmetric, in the sense that the expectation value of the trace of the holonomy becomes approximately zero.  In pure Yang-Mills theory the distribution becomes exactly center-symmetric for all $L$ larger than some critical value, while for one-flavor QCD the VEV of the trace of the holonomy only approaches zero as $L \to \infty$ when the quarks are light. In any case, we assume that the trace of the holonomy continues to be center-symmetric at small $L$.  Justifying this assumption relies on the introduction of an appropriate double-trace deformation or heavy adjoint fermions, and we assume that this has been done.

The color holonomy has the form $\Omega = \mathcal{P}e^{i \int dx_3 A_3}$, where $x_3$ is the compact direction of circumference $L$, and we assume that $L \ll \Lambda^{-1}$.  If $\langle \tr \Omega\rangle =0$, then (with an obvious gauge-fixing understood) $\langle A_3 \rangle \neq 0$, the eigenvalues of $A_3$ have parametrically small fluctuations thanks to the smallness of $L\Lambda$, and take the center-symmetric vacuum expectation value (VEV) 
$
\langle LA_3\rangle = \mathrm{diag}(-\pi/2,+\pi/2)
$.
If we integrate out Kaluza-Klein (KK) modes on $S^1$,  we obtain a 3D $SU(2)$ gauge theory.  In this 3D effective theory, $A_3$ acts like a {\it compact} adjoint Higgs field.   When the VEV of $A_3$ is center symmetric,  the color gauge group $SU(2)$  is reduced down  to $U(1)$, and theory Abelianizes at distances larger than $m_W^{-1} = NL/2\pi = L/\pi$.  So in fact the long-distance theory is 3D compact QED
\begin{align}
S_{\textrm{small L}} &= \frac{L}{4g^2}\int d^{3}x F_{\mu \nu}^2 + \textrm{matter} \,,
\end{align}
where $\mu, \nu = 0,1,2$ parametrize the non-compact directions, and $F_{\mu \nu}$ is the field strength associated with the unbroken $U(1)$ gauge group.  We will find useful to pass to the Abelian dual formulation in terms of  scalar $\sigma$, $F_{\mu \nu} = \frac{\lambda m_W}{4\pi^2} \epsilon_{\mu\nu\gamma}\partial^{\gamma} \sigma$, where $\lambda = g^2 N $ is the 't Hooft coupling.  Then the action of the 3D effective field theory takes the form
\begin{align}
S_{\textrm{small L}} 
&= \int d^{3}x \frac{\lambda m_W}{16\pi^3} (\partial_{\mu} \sigma)^2 + \textrm{matter} \,.
\end{align}

\subsubsection{The chiral limit $m_q = 0$} 
We take  
\begin{align}
\Psi(x_3+L) = e^{i \alpha} \Psi(x_3)
\end{align}
 as the fermion boundary condition.  The $\alpha$ twist in the boundary conditions can be thought of as a holonomy for a background gauge field associated to the $U(1)_Q$ symmetry of the theory. 
  The role of the twist $\alpha$ is to produce a parity-invariant mass term in the 3D effective theory,   
\begin{align}
 \int_{\R^3 \times S^1}   \, \,  \frac{\alpha}{L}  \;  \bar \Psi  \gamma_3 \Psi 
\end{align}   
This type of mass term is also often called a ``real mass term" in 3d language. 
 If we were to set $\alpha = 0$, corresponding to periodic boundary conditions, the long-distance theory would be  3D QED coupled to $n_f = 2$ flavors of massless Dirac fermions.  This EFT would be strongly-coupled on distances $\ell \gtrsim L/g^2$, where $g$ is the YM coupling.  This issue can be avoided by keeping $\alpha$ finite, because $\alpha$ produces the real mass term for the 3D fermions, as written above.  Consequently, so long as  $\alpha \gtrsim g^2$, our theory is weakly coupled at all distance scales, and can be treated reliably using semiclassical methods.  Thus there are no infrared divergences in perturbation theory.
 
Since our long-distance theory is 3D compact QED, it has finite-action monopole-instanton field configurations\cite{Polyakov:1976fu}.  In fact, because the theory is locally four-dimensional, the theory has \emph{two} types of monopole-instantons:  one is the standard BPS monopole-instanton which is present even in locally three-dimensional theories, while the other is the so-called Kaluza-Klein (KK) monopole instanton\cite{Lee:1997vp,Kraan:1998sn}.   We write the corresponding amplitudes as $\mathcal{M}_1,\mathcal{M}_2$.   These finite-action field configurations can be thought as the constituents of the 4d instanton $\mathcal{I}$ on the compactified geometry, $\mathcal{I} \sim \mathcal{M}_1 \mathcal{M}_2$.  Note that the monopole-instantons have no scale moduli, since they have a finite core size $\sim m_W^{-1}$.  Indeed, in this setting, a 4d BPST instanton with a large scale modulus $\rho \gg L$ looks like two widely separated $\mathcal{M}_1$, $\mathcal{M}_2$ events.  Along with our remarks concerning the gauge coupling, this implies that there are no infrared problems either perturbatively or non-perturbatively.

The two fermion zero modes of the 4d instanton  localize on one of the two types of monopole instantons, depending on the $U(1)_Q$ holonomy $e^{i\alpha}$.  For our choice of $\alpha$, we can take the fermion zero modes to be localized on $\mathcal{M}_1$.  To leading non-trivial order in the semiclassical expansion, the resulting dimensionless monopole operators take the form: 
\begin{align}
&{\cal M}_1 \sim  e^{-S_0} e^{i \sigma }   \psi_L \psi_R ,  \qquad {\cal M}_2 \sim  m_W^3 e^{-S_0}  e^{-i \sigma }, \cr
& \overline{\cal M}_1 \sim e^{-S_0} e^{-i \sigma }  \bar \psi_L \bar \psi_R ,  \qquad \overline {\cal M}_2 \sim   m_W^3 e^{-S_0}  e^{+i \sigma },  
\end{align}
where $S_0= \frac{1}{2}S_I= \frac{4\pi^2}{g^2}$  is the monopole-instanton action, which is half of the 4d-instanton action.\footnote{We have suppressed powers of $\lambda$ (which arise from summation over bosonic zero-modes) in the prefactors of the monopole-instanton amplitudes to reduce clutter in our expressions. These prefactors are identical for  $\mathcal{M}_{1}, \mathcal{M}_{2}$, and are not important for our discussion in this paper. }

To find the leading effect of the existence of monopole-instantons on the long-distance physics, we must sum over the dilute monopole-instanton ``gas"\cite{Polyakov:1976fu}.  At leading order, the monopole-instanton contributions to the partition function simply exponentiate, so that\cite{Unsal:2008ch}
\begin{align}
V(\sigma) &= - \left[\mathcal{M}_1 +  \overline{\mathcal{M}}_1 +  \mathcal{M}_2 +  \overline{\mathcal{M}}_2 + \cdots\right]  \\
&= - \left[  (e^{-S_0} e^{i \sigma }   \psi_L \psi_R + \mathrm{h.c.})   +  e^{-S_0} \cos(\sigma) + \cdots\right] \, .
\end{align}

The $\cos(\sigma)$ potential renders gauge fluctuations massive, with a mass $m_\sigma \sim m_W e^{-S_0/2}$.  The unique minimum of the potential is at $\sigma=0$, and this generates a  
``constituent quark mass" $m_{\rm constituent} \sim  m_W e^{-S_0}$. Note that this is exponentially larger than the contribution to the constituent quark mass from instantons, which is  $\sim m_W e^{-S_I} \sim m_W e^{-2S_0} $.

One can also introduce a topological $\theta$ angle.  But in the $m_q=0$ limit, it has no effect on the effective potential since it can be absorbed into fermion field by a field redefinition. 

\subsubsection{Finite $m_q \neq 0$}

\begin{figure}
\centering
\includegraphics[width=0.6\textwidth]{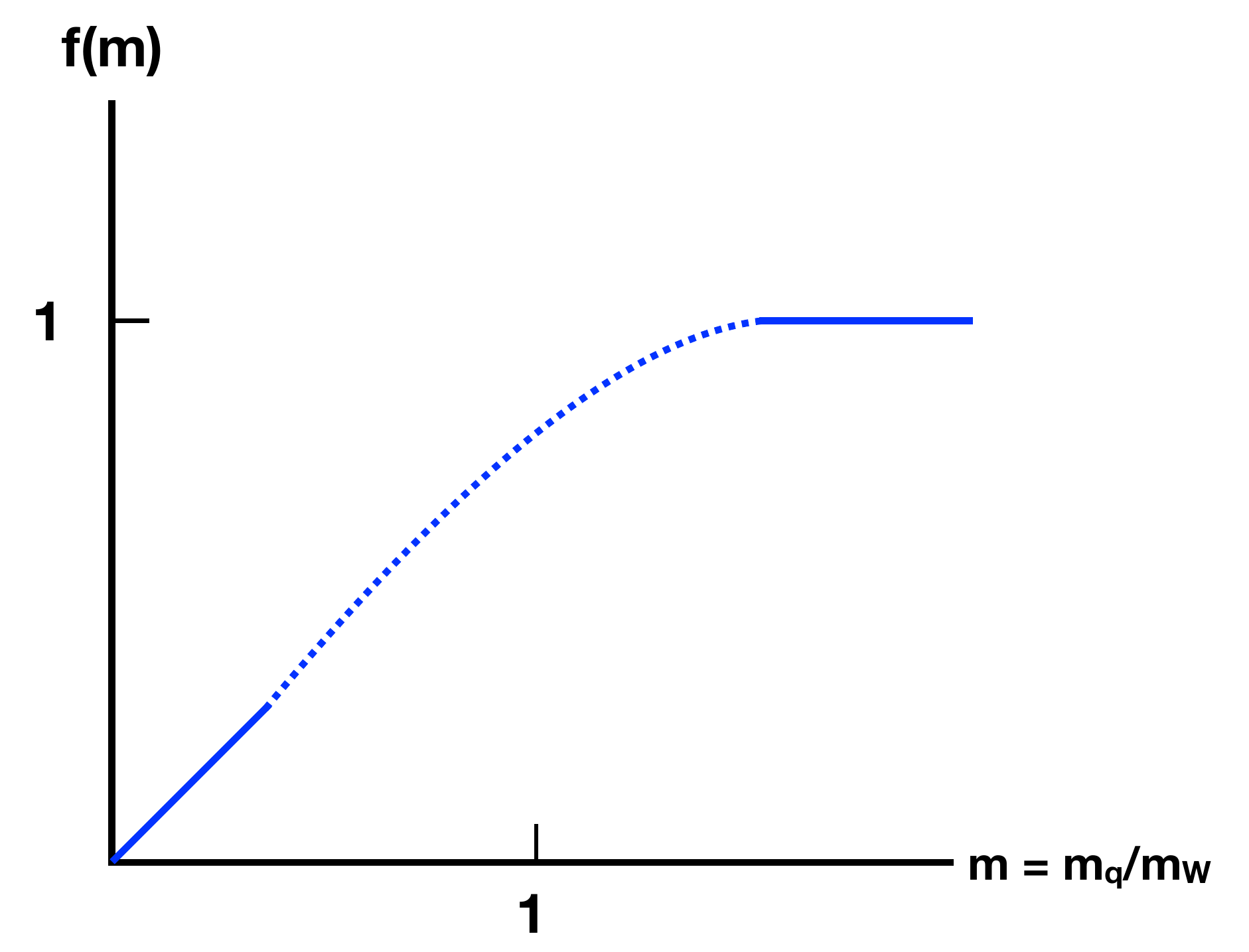}
\caption{A cartoon of the expected behavior of the massive fermion determinant $f(m)$ around a monopole-instanton as a function of the parameter $m = m_q/m_W$.  We assume that the quark boundary conditions are such that the fermion zero modes would localize on the monopole-instanton in the limit $m \to 0$, so that the small-$m$ behavior is $f(m) \sim m$, and assume a renormalization prescription such that the determinant approaches unity in the pure YM limit $m\to \infty$.  The solid parts of the curve summarize these limiting behaviors, while the dashed part is an interpolation, since the functional form of $f(m)$ has not yet been evaluated.
}
\label{fig:fm_sketch}
\end{figure}

Turning on a small quark mass $m_q$ lifts the fermionic zero modes, and renders the $\theta$ angle physical.  
The contribution of the monopole-instantons to the dual photon potential is now
\begin{align}
V(\sigma)= -  f(m) e^{-S_0}  \cos({\sigma +  {\theta \over 2}})  -  e^{-S_0} \cos(\sigma - {\theta \over 2}) -    f(m)   e^{-2S_0}  \cos({2 \sigma }) +  \ldots
\end{align}
where $m = m_q / m_W$ and  $e^{-2S_0}$ term is due to magnetic bion.\footnote{Actually, the bion potential term should have has various factors we have not shown which multiply $f(m)$, which involve $\log \lambda$ and powers of $\lambda$.  They could easily be restored, but have no important effect in the discussion, and we have dropped them to avoid cluttering the formulas. }   Here by $f(m)$ we mean the quark-mass-dependent prefactor of the monopole-instanton amplitude, normalized so that
$f(m) = \frac{\det(\Dslash\big|_{\rm{monopole}} +m_q)}{\lim_{m\to\infty} \det(\Dslash\big|_{\rm{monopole}} +m_q)} $,
where the functional determinants are evaluated in the background of the monopole-instanton.  
It is easy to determine  the behavior of $f(m)$ in two limiting cases:
\begin{align}
f(m)= \left\{\begin{array}{ll}c m + O(m^2)    \qquad &m \lesssim 1 \cr
 1+ O(1/m)  & m \rightarrow \infty  
 \end{array} \right.
 \label{extrapol}
\end{align} 
Here $c$ is an immaterial number of order one. We set it to unity to avoid clutter;  it is easy to restore it in the formulas below if one wishes.  We expect the cross-over value of $m_q$ setting the scale where $f(m)$ changes from scaling with $m_q$ to becoming $m_q$-independent,   to be $\sim m_W$, since this is the only dimensionful scale entering the monopole-instanton field profile. However, the functional form of $f(m)$ apart from the two limits in \eqref{extrapol}  has not yet been determined, and  might not have a closed-form expression, given analogous computations for BPST instantons\cite{Dunne:2004sx}.  A cartoon showing the expected shape of $f(m)$ is given in Fig.~\ref{fig:fm_sketch}.  

There are two crucial points to note about this potential.  Despite the fact that the magnetic charge of ${\cal M}_1$ and ${\cal M}_2$  is the same,  there is a phase difference between the induced operators, which is dictated by phase of $m_qe^{i \theta}$. The other point is that 
the magnitudes  $|  {\cal M}_1| $ and $|  {\cal M}_2|$ differ in an interesting way.  In modulus, $\mathcal{M}_1$ and $\mathcal{M}_2$ are very different for 
$m \lesssim 1$,  but become almost (but not quite) equal for $m > 1$.   So
\begin{align}
\arg {\cal M}_1 - \arg {\cal M}_2= \arg (m_q e^{i \theta} ) \,  \cr
|  {\cal M}_1| \neq  |  {\cal M}_2|,  \qquad    |m| \lesssim 1.
 \label{phases}
\end{align} 
Only in the  limit  $m \rightarrow \infty$ do the magnitudes of $\mathcal{M}_1$ and $\mathcal{M}_2$  become equal. 
 Both of these observations will 
 will play a crucial role in the physics at $\theta=\pi$.

\subsubsection{Large negative $m_q$}
When $|m_q| \gg m_W$, the theory reduces to deformed Yang-Mills on small $S^1 \times \R^3$,   which is  believed to be continuously connected to pure Yang-Mills on $\R^4$, because all gauge invariant order parameters have the same behavior at large and small $L$.   In this limit, the prefactors of the monopole operators become equal in absolute value, and the effective potential is given by 
\begin{align}
V(\sigma)= - m_W^3 \left[ e^{-S_0} \cos(\sigma +\theta/2) + e^{-S_0}  \cos({\sigma - \theta/2})   +  e^{-2S_0}  \cos({2 \sigma }) +  \ldots \right]
\end{align}
  The last term $\propto e^{-2S_0}$ is induced by magnetic bion events, which are quasi-saddle-points which become relevant at second order in the semiclassical expansion\cite{Unsal:2007jx,Unsal:2012zj}.  The first two terms dominate over the third term for $\theta \neq \pi$, but they cancel as $\theta \to \pi$, where the bion-induced term becomes dominant.  While the density of the monopole-instanton events of the two types are equal, their effects cancel due to destructive topological interference \cite{Unsal:2012zj}.  Indeed, at $\theta=\pi$, all odd charge monopole induced terms of the form $\cos({(2k+1) \sigma})$ are absent due to topological $\mathbb Z_2$ shift-symmetry, $\sigma \rightarrow \sigma+ \pi$.  But thanks to the contributions of the magnetic bions, the theory always has a mass gap.  For $\theta \neq \pi$, the ground state is gapped and unique.  But for $\theta = \pi$, there are two gapped vacua associated with CP breaking, 
   $\sigma = \pm \pi/2$, which are associated with CP breaking, since CP acts on $\sigma$ by $\sigma \to -\sigma + \pi/2$.  

Reference \cite{Gaiotto:2017yup} recently showed that there exists a mixed 't Hooft  anomaly between the 1-form center symmetry and the 0-form CP symmetry at $\theta=\pi$.  The anomaly strongly constrains the nature of the possible ground state of the theory at $\theta = \pi$.
 In particular, if (1) center symmetry is stable, and (2) the mass gap does not vanish, anomaly-matching implies that the CP must be broken.   But of course anomaly considerations cannot determine whether the two assumptions leading to CP breaking are in fact satisfied.  This requires information about the dynamics, either from  lattice simulations, or from some analytic approach.  But Euclidean lattice simulations suffer from a severe sign problem at $\theta = \pi$. 
 
This is  where semi-classical calculations enabled by adiabatic compactification are helpful.  Indeed, in center-stabilized YM theory, the semiclassical analysis presented in \cite{Unsal:2012zj} and reviewed above explicitly shows that CP is spontaneously broken at $\theta = \pi$.  If the center symmetry is stable for all $L$, which can be guaranteed by  an appropriate double-trace deformation at any $L$, and the mass gap remains finite for any finite $L$, which is expected at large $L$ and is shown analytically above for small $L$, then the two-fold degeneracy of the vacua due to spontaneous CP-breaking will persist for any $L$.

\subsubsection{Small negative $m_q$}
We now discuss what happens as $m_q$ approaches $0$ with $\theta = \pi$. It is convenient to do so using the shifted variable $\sigma' = \sigma + \theta/2$.  Then the potential reads
\begin{align}
V(\sigma')= -  f(m) e^{-S_0}  \cos(\sigma' +  \theta )  -  e^{-S_0} \cos(\sigma' ) -    f(m)   e^{-2S_0}  \cos(2 \sigma' +\theta) +  \ldots \, ,
\label{eq:YM_N2_shifted}
\end{align}
and at $\theta = \pi$ we obtain
\begin{align}
V(\sigma')=
 [f(m)-1] e^{-S_0} \cos(\sigma')  + f(m)  e^{-2S_0}  \cos({2 \sigma' }) +  \ldots \,.
\end{align}

\begin{figure}
\centering
\includegraphics[width=0.8\textwidth]{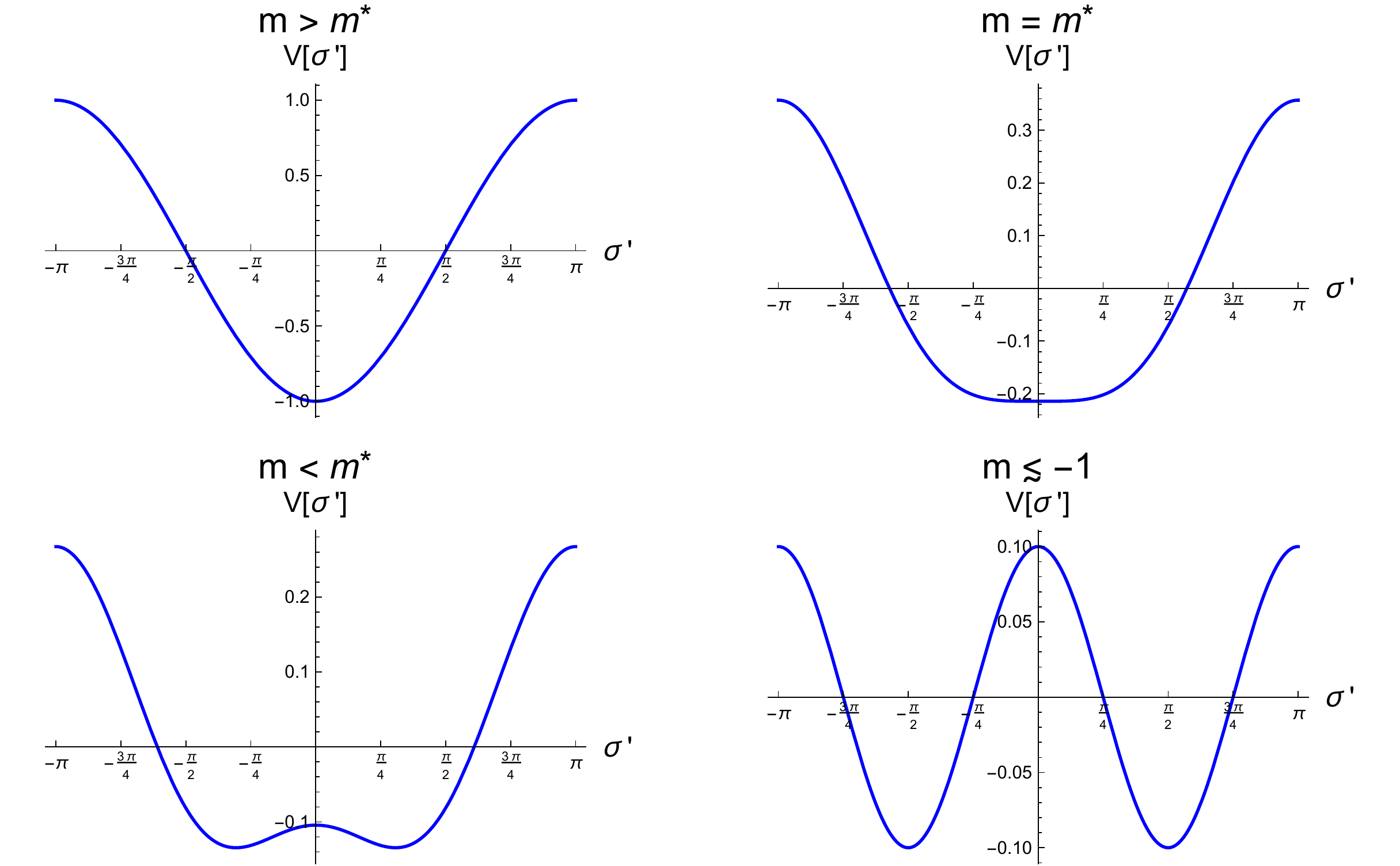}
\caption{Behavior of $V(\sigma')$ as a function of $m = m_q/m_W \in \mathbb{R}_{-}$, with $f(m)$ modeled as $f(m) \sim m$ for $|m| \lesssim |m^{*}|$.  The theory becomes gapless at $m = m^{*}$.}
\label{fig:SU2}
\end{figure}

As we have just seen, when $\theta = \pi$ and $m$ is large, there are two degenerate vacua and a corresponding first-order phase transition.   The two vacua are related by time-reversal (equivalently, CP) symmetry.  When $m$ is large and positive, the potential $V(\sigma')$ has a unique ground state, where $\langle \sigma' \rangle  = 0$.   As $m$ is decreased toward zero with fixed $\theta = \pi$, there is a second order critical point $m = m_{*}$ at which mass of the $\sigma'$ fluctuation vanishes, and that the theory becomes gapless.   Once $m < m_{*}$, there are again two degenerate vacua, and a corresponding first order phase transition.  These features are summarized in Fig.~\ref{fig:SU2}.

Given a smooth function $f(m)$ with the limiting behavior given by \eqref{extrapol}, there exists a value, $m_*$, at which the theory becomes gapless.  Indeed, expanding to 
\begin{align}
V(\sigma') =\half  \left[1-f(m)(1+ 4 e^{-S_0})  \right] \sigma'^2 
+\tfrac{1}{24} \left[-1 + f(m)(1 +16   e^{-S_0})\right] \sigma'^4 
+\mathcal{O}\left(\sigma'^6\right) \,.
\end{align}
So the mass of the $\sigma$ fluctuations vanishes when $m$ is such that
\begin{align} 
f(m) = 1 - 4 e^{-S_0} + O(e^{-2S_0} ) \,.
\end{align}
Given that $f(m)$ is a monotonic function with the limiting behaviors shown above, this equation will have one solution, $m = m^{*}$, where the theory has a gapless second-order critical point.

In this $N=2$ analysis, if we wish to get an explicit estimate of $m_{*}$, we must use a model for the precise shape of $f(m)$, since this shape is not currently known.  For instance, if we model $f(m) = m$, then we get $m^{*} \sim 1-4e^{-S_0}$.  But the existence of critical point is robust and model-independent, since it follows from the known limiting behaviors of the smooth function $f(m)$.  Moreover, as we are about to see, when $N\gg 1$, it is possible to compute the value of $m^{*}$ explicitly without making ad-hoc assumptions about the particular functional form of $f(m)$.

\subsection{Generic $N$}
For generic $N$ and to leading order in the semiclassical expansion, the low-energy EFT describing monopole instantons can be written as
\begin{align}
\int d^3 x \sum_{a=1}^{N}  \lambda m_W (\partial_{\mu} \sigma_{a} )^2 - m_W^3 e^{-S_0}&\bigg[ f(m) \cos(\sigma_1 - \sigma_2 +\theta) + \cos(\sigma_N - \sigma_1) \nonumber \\
&+ \sum_{a=2}^{N-1} \cos(\sigma_{a} - \sigma_{a+1})  \bigg] + \ldots \,.
\label{eq:generalNaction}
\end{align}
Here $f(m) = f(|m|)$.  In writing this expression, we have assumed that the approximate center symmetry (which becomes exact as $N\to \infty$ or $m_q \to \infty$) has been preserved at small $L$.  Then \eqref{eq:generalNaction} follows from essentially the same arguments as above.  The gauge group is Higgsed from $SU(N)$ to $U(1)^{N-1}$, and the gauge coupling stops running at the scale $m_W = 2\pi/(NL)$.   So in the Abelian dual description there are now $N-1$ dual scalar fields $\sigma_i$.  In writing \eqref{eq:generalNaction}, however, we chose to add one extra dual scalar field, as if we had started with a $U(N)$ gauge theory rather than an $SU(N)$ gauge theory, because this allows us to write many expressions in a simpler form.  As will be clear below, this extra scalar decouples exactly from the $N-1$ physical fields. There are now $N$ types of monopole-instantons, with $N-1$ of them being standard 't Hooft-Polyakov monopoles, while the $N$th one is the KK monopole-instanton.  The associated amplitudes are
\begin{align}
\mathcal{M}_{1} &= m_W^3 f(m) e^{-S_0} e^{+i(\sigma_1 - \sigma_{2})} e^{i\theta},  \\
\mathcal{M}_{i} &= m_W^3 e^{-S_0} e^{+i(\sigma_i - \sigma_{i+1})} \, , \; i =2, \ldots, N, 
\end{align}
where  $N+1 \equiv 1$. Here,  we have assumed that the $U(1)_Q$ twist angle $\alpha$ is such that the fermion zero modes localize onto $\mathcal{M}_1$ in the chiral limit. We have also exploited the freedom to perform field redefinitions to put all of the $\theta$ dependence onto $\mathcal{M}_1$.  (Section~\ref{sec:transmutation} contains an in-depth discussions of such field redefinitions, and the related phenomenon of $\theta$-dependence transmutation as $m_q$ goes from small to large.)
These $N$ species of monopole-instantons are again the constituents of the familiar 4d BPST instanton, 
\begin{align}
\mathcal{I} \sim \mathcal{M}_1 \cdots \mathcal{M}_N, \qquad S_0= \frac{S_I}{N}= \frac{8 \pi^2}{g^2N}.
\end{align}
Evaluating the sum over the dilute monopole-instanton gas to leading non-trivial order yields \eqref{eq:generalNaction}.

For $N_F=1$ and finite $N$, there is no exact (axial) chiral symmetry in the $m_q=0$  limit.   For generic values of $m = m_q/m_W$, we will now see that all physical modes  have mass square proportional to  $e^{-S_0}$,   but  one linear combination of the modes in $\vec{\sigma}$ describes $\eta'$ meson.  However,  along a ray on the complex mass plane 
$m \in [-\infty, m_c e^{i \pi})$, CP symmetry is spontaneously broken.  This CP-breaking ray of first-order transitions ends at $m_c e^{i \pi}$ with a  second order critical point. 

\begin{figure}
\centering
\includegraphics[width=1.\textwidth]{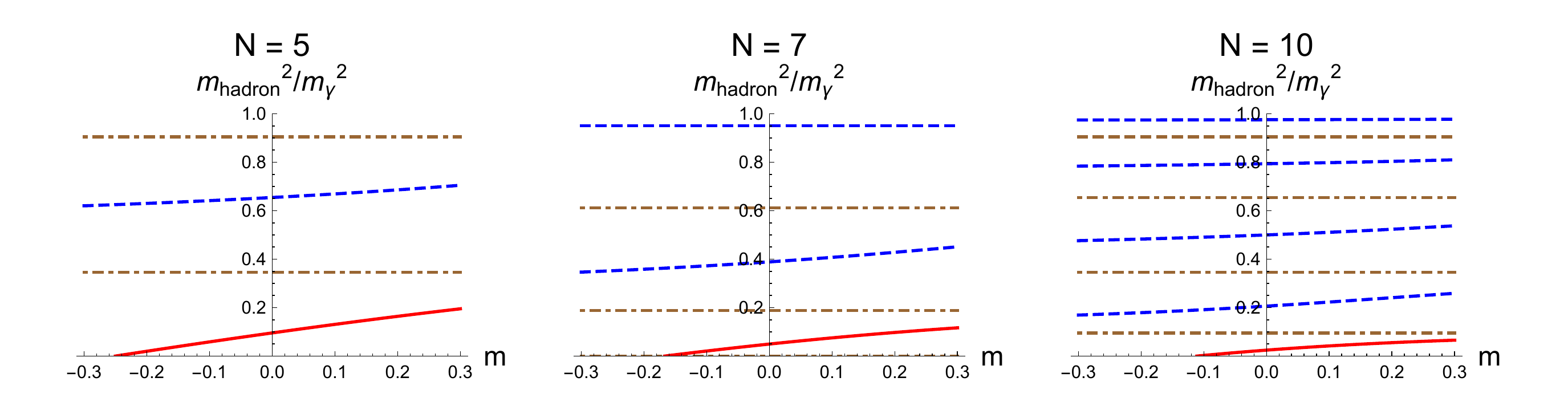}
\caption{A plot of the squared masses of the four lightest pseudoscalar hadrons in adiabatically-compactified $N_F=1$ QCD with $N=5,7,10$ colors for small quark mass $m_q$, plotted as a function of $m = m_q/m_W$, with the sign of $m$ corresponding to whether $\theta = 0$ or $\theta = \pi$.  The lightest mode, indicated by the solid red line, becomes gapless at smaller negative $m$ as $N$ increases.  So this can be identified as the $\eta'$ meson.  The dashed blue curves correspond to the modes whose masses are independent of $m$ at leading order in the semiclassical expansion, while the dot-dashed brown curves are pseudoscalar modes with masses that do depend on $m$.  The $m$ dependence of these modes decreases as $N$ is increased.}
\label{fig:etaprime}
\end{figure}

To see this, let us set $\theta = 0$ and $\theta = \pi$ (so that we study the theory along the real axis in the complex-mass plane), and suppose that the minimum of the potential is at $\vec{\sigma} = (0,\ldots, 0)$.   This assumption will be valid provided $m_q> m^{*}$.    Expanding around this minimum, the mass matrix is
\begin{align}
M = \begin{pmatrix}
1 \pm f(m) & \mp f(m)  &   &  &      & & -1\\
\mp f(m)    & 1 \pm f(m)  & -1 &        & & & \\
    &    -1 &  2 &  -1    &  && \\
      &        &  -1 &  2    & & & \\
        &        &   &     &  \ddots&& \\
      &        &   &        && 2 & -1 \\
-1      &        &   &     && -1 & 2 
\end{pmatrix}
\end{align}
where the upper and lower signs correspond to the effect of setting $\theta = 0$ and $\theta = \pi$ respectively.
For  $m > m^{*}$, the spectrum of the $M$ operator is positive definite, corresponding to the hadron spectrum of the small-circle theory. 
However, when $\theta = \pi$, there is a point 
\begin{align}
f(m) = f(m_{*}) = +1/(N-1) \,, 
\label{eq:generalN_mass_condition}
\end{align}
 there is an eigenvector $v_{\eta'}$ such that $M v_{\eta'} = 0$, which takes the form
\begin{align}
v_{\eta'} = \left(-\tfrac{N-1}{2}, \tfrac{N-1}{2},\tfrac{N-1}{2}-1, \ldots, \tfrac{N-1}{2}-(N-2)\right) \, .
\end{align}
We identify the resulting gapless mode with the $\eta'$ meson.\footnote{Generically, $M$ has one other vanishing eigenvalue for any $m$, with eigenvector $v_{N} = N^{-1/2} (1, 1, \cdots, 1)$.  The corresponding gapless excitation can  be shown to be exactly free.  In fact, this excitation is unphysical, and arises from our use of an $N$-dimensional basis in writing the $\mathfrak{su}(N)$ roots and dual photon vectors.}   This identification is supported by the fact that this mode is the lightest pseudoscalar excitation of the theory at $m_q = 0$, as we illustrate in Fig.~\ref{fig:etaprime}, as well as our discussion in Section~\ref{sec:generic}, where we explain how this mode enters the chiral Lagrangian.    Note that the $\eta'$ mode is not free already to leading order in the semiclassical expansion:  the potential \eqref{eq:generalNaction} generates an $\eta'^4$ interaction term with coefficient $\sim e^{-S_0}$.

It is also interesting to note that when $N$ is large, we can use the known behavior of $f(m)$ for small $m$ to conclude that, in terms of the complexified quark mass $\mathfrak{m}_q = m_q e^{i\theta}$, when $\theta = \pi$ we get
\begin{align}
\frac{\mathfrak{m}^{*}_q}{m_W} = \frac{-1}{N-1} + O(e^{-S_0})
\end{align}
So at large $N$, the location of the fixed point can be self-consistently calculated using only the known properties of $f(m)$.   Moreover, in the $N \to \infty $ limit,  the $\eta'$ meson becomes gapless at $m_q=0$, just as one would expect from the fact that an unbroken $U(1)_A$ symmetry must emerge at $m_q = 0$ in the large $N$ limit.

\section{How can a  3d Coulomb gas ever fail to generate screening?}
\label{sec:Coulomb}
The gauge theory on  $\R^3 \times S^1$ maps to a (magnetic)  Coulomb gas. A Coulomb gas in $D= 3$ is known to have a finite Debye length.  For example, in the Polyakov model, this finite Debye length is the inverse mass gap of the system.  
There is a rigorous renormalization group argument due to Kosterlitz \cite{0022-3719-10-19-011} which also shows that the  Coulomb gas is always in a screening phase for  $D= 3$. In this case, the $\cos( \sigma)$ term that generates the mass gap is relevant. As a result, at long distance one obtains Debye screening.

The Kosterlitz argument is also valid for systems which are effectively generalized Coulomb gases, as discussed in \cite{Fradkin:1991nr}. 
In some systems, in the  monopole-gas
representation, the action possesses an imaginary part, coming either from (lattice/cut-off  scale) Berry phases or 
 a topological theta angle.  Examples include the  valence-bond-solid (VBS) phase of quantum anti-ferromagnets \cite{Read:1990zza, Fradkin:1991nr}   and  center-stabilized Yang-Mills at $\theta = \pi$ \cite{Unsal:2012zj}.  In these cases, in order to derive the long-distance  effective field theory, one can do a succession of ``coarse-graining" steps, as is common in statistical mechanics treatments of renormalization. 
This process gets rid of $e^{-S_0} e^{i\sigma} e^{i \Theta_i}$ type operators,  due to destructive interference induced by Berry phases or the topological interference due to $\Theta$-angles among monopoles with the same magnetic charge, but differing  phases.  
Namely, one has a relation of the form 
\begin{align}
 \sum_{j=1}^{K} {\cal M}_j  = \sum_{j=1}^{K}  e^{-S_0} e^{i\sigma} e^{i \Theta_j} =0  \qquad \Theta_j= \frac{2 \pi j }{K}
 \label{cancel}
 \end{align} 
In the statistical-mechanics examples, such relations are possible because the magnitudes of the distinct monopoles amplitudes with identical magnetic charge are forced to be identical by exact lattice symmetries.  In Yang-Mills theory, the equality of magnitudes of monopole-instanton amplitudes is guaranteed by center symmetry.  However, there are always some monopoles with higher charge which do not carry Berry phases, or any dependence on the topological $\theta$ angle.  Such monopoles survive the coarse-graining procedure. In $SU(2)$ gauge theory on $\mathbb{R}^3\times S^1$, these are the magnetic bions, which have magnetic charge $2$ and vanishing topological charge, while in the VBS phase of quantum anti-ferromagnets, the surviving contributions are from certain multi-charge monopoles. 
Generically, the surviving operators induce a potential for the dual photon $\sigma$ of the form 
\begin{align}
V(\sigma) \sim e^{-nS_0} \cos(n \sigma)
\end{align}
for some $n>1$.  Then the Kosterlitz argument goes through, because this potential always includes relevant terms.
The fluctuations of the dual photon field  $\sigma$  become suppressed,
since the relevant  cosine operator  pins the fluctuations down to the bottom of one of its minima. 
And hence, one ``always" obtains a finite screening length for generalized Coulomb gas as well. The arguments described above have lead to intuition in the literature that Coulomb gases in three dimensions are always gapped.   

The only historically known exception appeared in the work of Affleck, Harvey, and Witten\cite{Affleck:1982as}, which added massless adjoint fermions to the non-Abelian gauge theory UV completion of a compact QED system, and observed that this rendered the system gapless.   But in the case considered in \cite{Affleck:1982as}, the system is not simply a magnetic Coulomb gas with bare interactions of the form $\sim 1/r$, because the exchange of fermion zero modes between monopoles immediately produces a logarithmic contribution to the potential $\sim \log r$.   As a result, this case does not fit into the paradigm discussed above.

In  light of the above discussion, one may wonder  how can one ever obtain a Coulomb gas which can be gapless.  If a Coulomb gas with infinite correlation length can exist, where is the flaw in the above classic arguments? 

The answer of course is already present in the analysis of the previous section, both for $N=2$ and $N>2$.  Let us focus on $N=2$.    We have a Coulomb gas of magnetic monopoles 
at $\theta = \pi$, and there is a relative $\pi$ phase between the two distinct monopoles of the same magnetic charge. If there were to be a precise symmetry between monopole ${\cal M}_1$ and  ${\cal M}_2$, their contributions would cancel each other out and leave room for contributions from higher charge operators, such as 
$e^{-2S_0} \cos(2 \sigma)$. However, in our system, the contributions of ${\cal M}_1$ and  ${\cal M}_2$ to the effective potential do not cancel each other out exactly, because the symmetry relating them --- center symmetry --- is \emph{explicitly} broken.  The amount by which these leading-order contributions fail to cancel can be dialed by changing the fermion mass parameter $m$.  In particular, the sign of the leading-order contribution to the mass of the dual photon can be arranged to be negative.   Then one can tune the negative leading-order contribution against the higher-order contributions, in such a way that the coefficient of $\sigma^2$ vanishes exactly.  Consequently, one finds that there is a second-order critical point where the theory is gapless.  Note that here there is no shift symmetry for $\sigma$, and $V(\sigma) \neq 0 $ at the critical point.   

It would be very interesting to reexamine the physics of VBS phases of quantum anti-ferromagnets discussed in \cite{Read:1990zza} in view of the above remarks.  In \cite{Read:1990zza}, there are lattice symmetries which ensure the equality of the magnitude of the unit-charge monopole amplitudes.  It might be possible to explicitly break these lattice symmetries in such a way that, as a function of the perturbation, the system can be tuned to a gapless critical point.  This would be analogous to the critical point in one-flavor 4d QCD as a function of $m$, where the relevant broken symmetry is center symmetry. 

\section{Generic  $N_F, N$}
\label{sec:generic}
We now let $N > 2$, and consider what happens when $N_F \le N$, with the assumption that the $N_F$ quark flavors have a common mass $m_q$, so that the global symmetry of the theory on $\mathbb{R}^4$ is
\begin{align}
G_{m_{q} \neq 0,\, \mathbb{R}^4} =\frac{SU(N_F)_V \times U(1)_Q}{\mathbb{Z}_{N_F} \times \mathbb{Z}_{N}} .
\label{eq:QCD_sym}
\end{align}
Here the $\mathbb{Z}_{N_F}$ and $\mathbb{Z}_{N}$ factors in the quotient act by $\psi_{a} \to e^{i 2\pi/N_{F}} \psi_{a}$, $\psi_{a} \to e^{i 2\pi/N} \psi_{a}$, and come from the action of center elements of $SU(N_F)_V$ and $SU(N)$ respectively. In the chiral limit $m_q \to 0$ the anomaly-free internal global symmetry is enhanced to
\begin{align}
G_{m_{q} = 0,\, \mathbb{R}^4} = \frac{SU(N_F)_L \times SU(N_F)_R \times U(1)_Q}{\mathbb{Z}_{N_F} \times \mathbb{Z}_{N}}\, ,
\end{align}
which breaks spontaneously breaks to $G_{m_{q} \neq 0,\, \mathbb{R}^4}$.

We compactify the theory on $S^1$ with the flavor-twisted boundary conditions described in e.g. \cite{Cherman:2017tey,
    Iritani:2015ara}
    \begin{align}
\Psi_a(x_3+L) = e^{i \alpha} e^{2\pi i a/N_F} \Psi_a(x_3) \,, \; \;\;  a = 1, 2, \ldots, N_F \, .
\label{eq:TBCs}
\end{align}
When $d = \mathrm{gcd}(N,N_F) >1$ these boundary conditions preserve a $\mathbb{Z}_d$ subgroup of $\mathbb{Z}_N \times \mathbb{Z}^{\prime}_{N_F}$, where $\mathbb{Z}_N$ is the global center symmetry group of $SU(N)$ gauge theory, while $\mathbb{Z}^{\prime}_{N_F}$ is the group of cyclic flavor permutations, which act by $\psi_{a} \to \psi_{a+1 \, \textrm{mod}\, N_F}$.   
Following \cite{Cherman:2017tey} we refer to $\mathbb{Z}_d$ as the color-flavor-center (CFC) symmetry.~\footnote{In the  setting where $N = N_F$, the same boundary conditions were discussed in \cite{Kouno:2012zz,
    Sakai:2012ika,
    Kouno:2013zr,
    Kouno:2013mma,
    Iritani:2015ara,
    Kouno:2015sja,
    Hirakida:2016rqd,Larsen:2016fvs,
    Hirakida:2017bye}.  In these references the construction is interpreted as defining a ``QCD-like" theory termed ``$\mathbb{Z}_3$ QCD". In contrast, in \cite{Cherman:2017tey} and here, we interpret the CFC symmetry as a bona fide symmetry of QCD itself.  As pointed out in \cite{Cherman:2017tey}, this symmetry has interesting order parameters whose behavior depends non-trivially on the parameters of the theory.}    Its order parameters include Polyakov loops wrapping the compactified $x_3$ direction, as well as some local operators built out of the quark bilinears \cite{Cherman:2017tey}. 
    
Note that the $\mathbb{Z}_{N}$ symmetry relevant to the CFC symmetry $\mathbb{Z}_d$ is different from the action of the $\mathbb{Z}_N$ factor in \eqref{eq:QCD_sym}.  The $\mathbb{Z}_N$ transformations relevant to the CFC symmetry are related to the \emph{global} center symmetry of pure YM theory arising from gauge transformations which are aperiodic by an element of the center, while the $\mathbb{Z}_N$ in \eqref{eq:QCD_sym} comes from the action of standard periodic gauge transformations. 

The boundary conditions in \eqref{eq:TBCs}  break $G_{\mathbb{R}^4}$ to 
\begin{align}
G_{S^1} = \begin{cases} \frac{U(1)^{N_F -1}_L \times U(1)^{N_F -1}_R  \times U(1)_Q}{\mathbb{Z}_{N_F} \times \,\,\Z_N } \times S_{N_F}\times \Z_{2N_F}\,, & m_q = 0 \\
 \frac{U(1)^{N_F -1}_V \times U(1)_Q}{\mathbb{Z}_{N_F} \times\,\, \Z_N} \times S_{N_F}\, ,  & m_q \neq 0 \, .
\end{cases}
\label{eq:max-ab}
\end{align}
where  $S_{N_F}$ is the Weyl group of $SU(N_F)$, which is the group of permutations of the $N_F$ quark flavors, while $\Z_{2N_F}$ is the non-anomalous remnant of the classical $U(1)_A$ symmetry acting as $\Psi_a \to e^{i \alpha \gamma_5} \Psi_a$.  The cyclic flavor permutation group $\mathbb{Z}^{\prime}_{N_F}$ is a subgroup of $S_{N_F}$. 
 In any case, the flavor-center-symmetric boundary conditions give the  charged Nambu-Goldstone bosons masses $\sim 1/L$, but the neutral Nambu-Goldstone bosons remainmassless when $m_q = 0$.   
 Lattice simulations and continuity arguments in $m_q$ imply that $\mathbb{Z}_d$ CFC symmetry is not spontaneously broken at large $L$\cite{Iritani:2015ara} for any $m_q$, provided that $N_F/N$ is not large enough to put the theory into the conformal window\cite{Cherman:2017tey}. 
 The CFC symmetry can be preserved at small $L$ either by adding heavy adjoint fermions or certain double-trace terms, just as in pure YM theory\cite{Cherman:2016hcd}, and we assume that this has been done from here onward.

Let us now discuss the behavior of the theory at small $L$.  When $m_q = 0$, a 4D BPST instanton has $2N_F$ zero modes.  These $2N_F$ zero modes are distributed over the $N$ monopole-instantons in a manner which is dictated by the quark boundary conditions\cite{Nye:2000eg,Poppitz:2008hr}. In particular, with the flavor-twisted  boundary conditions in \eqref{eq:TBCs}, which are equivalent to turning on a flavor-center-symmetric holonomy for a background $SU(N_F)$ gauge field, the zero modes are distributed among the monopoles as evenly as possible \cite{Cherman:2016hcd}.

It is sensible to split the leading monopole-instantons  with action $S_0 = \frac{8 \pi^2}{g^2N}$ into two sets.  One set, 
$\mathcal{S}_1$, consists of $N_F$ monopole-instantons which carry two fermionic zero modes each.  The other set, 
$\mathcal{S}_2$, consists of $N - N_F$ monopole-instantons which do not carry any fermionic zero modes.  The associated amplitudes take the form
\begin{align}
\label{monopoles-2}
{\cal M}_{i} &=    e^{-S_0}  e^{i \vec\alpha_i \cdot \vec\sigma } 
        \psi_{Li} \psi_{Ri}\,,
   \;\; \;\; \qquad i \in  \, {\cal S}_1  ,  \\
{\cal M}_{k} &=  e^{-S_0}   e^{i \vec\alpha_k \cdot \vec\sigma }, 
                                  \qquad   \qquad \qquad  k  \in  \, {\cal S}_2  \, . 
\end{align}
The monopole operators in  ${\cal S}_1$   play a major role in the low energy dynamics, and especially in the emergence of the  chiral Lagrangian.  

This set-up actually provides a  microscopic semi-classical  derivation of chiral Lagrangian on 
$\R^3 \times S^1$  \cite{Cherman:2016hcd}, which we now briefly review. In the absence of the monopole operators, the dual photon has  $[U(1)_{ J}]^{N -1}$ topological shift symmetry.  In general there is no such symmetry in the original 4D gauge theory, so we would expect it to be broken by non-perturbatively induced interactions.  And indeed, the operators in  ${\cal S}_2$ 
break  $[U(1)_{ J}]^{N -1}$ explicitly to a subgroup as they induce a potential for certain modes. However, the effect of the operators in  ${\cal S}_1$ is more subtle. To see this, 
note that the flavor-twisted boundary preserve a $U(1)_A^{N_F-1}$ anomaly-free chiral symmetry, as seen in \eqref{eq:max-ab}. As a result, the monopole operators in $\mathcal{S}_1$ must be invariant under this symmetry.  But this is impossible unless the dual scalars transform  by shifts under $U(1)_A^{N_F-1}$, since     $\psi_{Li} \psi_{Ri}$ transforms under the chiral symmetry.  In particular, for the set of monopole-instantons in ${\cal S}_1$, we have:
 \begin{align}
\begin{array}{ccc}
    \psi_{Lk} \psi_{Rk}  & \rightarrow &
   e^{i \epsilon_k }  \psi_{Lk} \psi_{Rk}  ,  \\[0.2cm]
e^{i \vec\alpha_k  \cdot \vec\sigma  } &\rightarrow & 
   e^{-i \epsilon_k }    e^{i\vec\alpha_k  \cdot \vec\sigma} .
\end{array} 
\end{align} 
The  $N_F$ parameters $\epsilon_k$ obey one constraint,  $\sum_{k=1}^{N_F} \epsilon_k
= 0$, which can be realized as $\epsilon_k =  \vec{\alpha}_k \cdot \vec{\epsilon}$ 
where $\vec \alpha_k$ are the simple roots of the Lie algebra $\mathfrak{su}(N_F)$.    This implies that the $[U(1)_A]^{N_F-1}$ symmetry gets intertwined with the $[U(1)_{ J}]^{N_F -1}$  part of the would-be topological shift symmetry of the dual  photons.  This happens in such a way that their diagonal combination is an exact symmetry of the theory in the $m_q \to 0 $ limit.   

Thus we find that $N_F-1$ dual photons transform by shifts under a continuous chiral symmetry\cite{Cherman:2016hcd}.  Therefore they cannot develop a mass either perturbatively or non-perturbatively.  Moreover, at any given point on the vacuum manifold,  $[U(1)_{A}]^{N_F-1}$ symmetry is spontaneously broken. The vacuum alignment relevant to the chiral limit can be worked in the same way it is always done in chiral perturbation theory: after finding the vacuum at finite positive $m_q$, which is $\vec{\sigma} = 0$, one can expand around this point and take the $m_q \to 0$ limit to find the spectrum.  The result is that the $N_F-1$ gapless dual photons are the interpolating fields for the gapless neutral Nambu-Goldstone bosons expected from the symmetries of the compactified theory.    
The remaining  $N-N_F$ dual photons acquire a non-perturbative mass  of order $ m_W e^{-S_0/2}$, as in pure center-stabilized YM theory at small $S^1$.\footnote{It is not currently known how to ensure adiabatic continuity when $N_F>N$.}

Now, when we introduce a bare mass term in the QCD action, the $\theta$ angle becomes physical.  One can absorb the theta angle into the mass term by a chiral rotation to yield ${m_q (\bar{\psi}_{Lk} \psi_{Rk}) } \rightarrow {m_q  e^{i \theta/N_F} (\bar{\psi}_{Lk} \psi_{Rk}) }$. Let us suppose that the quark mass is small, $m_q \ll m_W$, so that we can write the dependence of the monopole operators on $m_q$ explicitly.    Then, soaking up the fermion zero modes with mass operator, 
 the monopole operators are modified into 
\begin{align}
\label{monopoles-3}
{\cal M}_{i} &=    e^{-S_0}  e^{i \vec\alpha_i \cdot \vec\sigma } 
      m e^{i \theta/N_F}  , \qquad i \in  \, {\cal S}_1  ,  \\
{\cal M}_{k} &=  e^{-S_0}   e^{i \vec\alpha_k \cdot \vec\sigma }   , 
                                  \qquad \qquad   \;\;\;\,\,  k  \in  \, {\cal S}_2  \, . 
\end{align}
where $m= m_q/m_W$. 

The effective potential induced by monopole-events is now
\begin{align}
V(\vec \sigma)= -  e^{-S_0}  f(m) \sum_{ i \in  \, {\cal S}_1}  \cos( \vec\alpha_i \cdot \vec\sigma + \theta/N_F ) -  e^{-S_0}   \sum_{ i \in  \, {\cal S}_2}  \cos( \vec\alpha_i \cdot \vec\sigma ) \,, \; N_F \neq 0
\label{Master}
\end{align}
This action is invariant under the global CFC symmetry $\Z_{\gcd(N_F, N)}$. We have already analyzed this action for $N_F=1$ in a previous section, so now we focus on $N_F>1$. 

First, for $ 2 \leq N_F \leq N-1$, $N_F-1$ modes are gapless in the  $m_q=0$ limit.  So $m_q = 0$ is the critical point of the theory, as one would expect.  These gapless modes are the Nambu-Goldstone  bosons associated to the spontaneously broken abelianized chiral symmetry. When $m \ll 1$, these modes have mass squares  proportional to  $m e^{-S_0}$, i.e, they are light compared to $e^{-S_0}$.    There is also another set of $N-N_F$ heavier modes whose mass squared is  $e^{-S_0}$, as well as other heavy modes with masses $\sim m_W$ we have not explicitly included in our 3D effective field theory. (They are discussed in detail in \cite{Aitken:2017ayq}.)  Note that at the critical point $m_q = 0$, the $\eta'$ meson is not massless, except in the limit $N\to\infty$ with fixed $N_F$. 

For $m_qe^{i \pi} \leq 0$, the effective theory has  two CP breaking vacua for  $ 2 \leq N_F \leq N-1$.  In  \cite{Gaiotto:2017tne}, it is shown that there is a mixed  anomaly between $SU(N_F)_V$ and CP symmetry when $d = \mathrm{gcd}(N, N_F)>1$.  In our case, there is an $SU(N_F)_V$ symmetry at short distances $\ell \ll 1/(N_{F}L)$, but it is broken to its maximal torus by the boundary conditions.  But there still exists a mixed anomaly between $U(1)^{N_F-1}$ and CP symmetry  when  $d>1$, similarly to the easy axis model \cite{Komargodski:2017smk}. Our results are consistent with the  constraints of this anomaly.  As we mentioned earlier, the anomaly does not determine the vacuum structure, but in this case the dynamical information encoded in \eqref{Master} implies that, of the several choices for the vacuum structure allowed by the anomaly, the theory chooses to have broken CP symmetry.

When $  N_F =N$, $N_F-1$ modes are gapless in the $m_q=0$ limit. In this case, in our effective Lagrangian based on dual photons,  there are no degrees of freedom corresponding to an $\eta'$ mode.  Of course, when $N_F = N$, there is no reason to expect the $\eta'$ mode to be parametrically light either at finite $N$ or large $N$.  Indeed, the full theory contains many more degrees of freedom beyond the dual photons, such as the fields corresponding to W-bosons and heavy quark modes.  These heavier modes have their own very rich dynamics discussed in \cite{Aitken:2017ayq}.  We expect that the $\eta'$ can be identified with the lightest isoscalar parity-odd heavy-mode bound state in the spectrum discussed in \cite{Aitken:2017ayq}, but have not explored this quantitatively.
 
It is important to note that the truncation of the monopole operators to the set ${\cal S}_1$ is equivalent to the reduction of our effective theory down to chiral Lagrangian.  
However, keeping heavier modes has its advantages. In particular, in \cite{Gaiotto:2017tne}, the  vacuum structure remains undetermined at leading order in chiral Lagrangian for $N_F=2$, as the associated effective potential term vanishes identically.  It is posited that sub-leading terms in the chiral Lagrangian renders vacuum two-fold degenerate. Moreover, in exploring the vacuum structure at generic $N_F$, \cite{Gaiotto:2017tne} assumes that $SU(N_F)_V$ symmetry is not broken spontaneously at $\theta = \pi$.  There is no known systematic argument ensuring this, since the Vafa-Witten theorem\cite{Vafa:1983tf} does not apply at $\theta = \pi$. In our case, since our effective theory has both light and heavy modes, the leading order in semi-classics actually suffice to prove that $CP$ is broken spontaneously for all $N_F \leq N$, while the vector-like part of flavor symmetry is not spontaneously broken.

To see the connection of \eqref{Master} to chiral Lagrangian,  construct the matrix $U$: 
\begin{align}
U = {\rm Diag} ({e^{i \gamma_i}  ),   \qquad \gamma_i=  \vec\alpha_i \cdot \vec\sigma, \qquad    i \in {\cal S}_1}  \, ,
\end{align}
 which is an  $N_F \times N_F$ diagonal matrix. Since ${\rm Det} U= e^{i  \sum_{   i \in {\cal S}_1}  \gamma_i}$, we can define 
\begin{align} 
\hat \gamma_i= \gamma_i - \frac{1}{N_F} \sum_{ i \in {\cal S}_1 } \gamma_i \equiv \gamma_i - \frac{\eta'}{N_F} \,
\end{align} 
The identification  $\eta'=  \sum_{ i \in {\cal S}_1 } \gamma_i$ will be shown later. 
Then, one can write $U$ as 
\begin{align} 
U  = e^{i \frac{\eta'}{N_F}} \tilde U \, ,
\end{align}
where $\tilde{U}$ has unit determinant.  
 
 Next,  note that we can also
assemble the ``heavy part" into an $(N-N_F) \times (N -N_F)$ square matrix.  Let 
\begin{align}
\Sigma= {\rm Diag} ({e^{i \gamma_i}  ),  \qquad  i \in {\cal S}_2}.  
\end{align}
 Clearly, ${\rm Det} \Sigma= e^{i  \sum_{   i \in {\cal S}_2}  \gamma_i}$, which can be written as  $\Sigma=  e^{i \delta} \tilde \Sigma$, where $\tilde \Sigma$ is special unitary. Since the direct sum  of $U$ and $\Sigma$ is also special unitary by construction,  $\delta= - \frac{\eta'}{N - N_F}$, resulting in 
 \begin{align} 
\Sigma=  e^{-i\frac{\eta'}{N - N_F} } \tilde \Sigma \, ,
\end{align} 
 Therefore,   an exact re-writing of monopole induced effective Lagrangian \eqref{Master} is 
\begin{align}
V& = -  e^{-S_0}  f(m)  e^{i \theta/N_F  + i \eta' /N_F} \sum_{ i \in  \, {\cal S}_1}  e^{i  \hat \gamma_i}  
 -  e^{-S_0}  e^{- i \eta' /(N-N_F)} \sum_{ i \in  \, {\cal S}_2}  e^{i  \hat \gamma_i}  + c.c. 
 \cr \cr
& = -  e^{-S_0}  f(m)  e^{i \theta/N_F  + i \eta' /N_F} \tr \tilde{U}   -  e^{-S_0}  e^{- i \eta' /(N-N_F)}   \tr \tilde{\Sigma}  + c.c.
\label{Master-3}
\end{align}
Note in particular that the matrix containing the light fields, $U$, has exactly the form one would expect from the standard unitary field appearing in a chiral Lagrangian, and its potential $\sim \tr( U + U^{\dag})$ is exactly of the form expected from the chiral Lagrangian.

Expanding the action associated to \eqref{Master-3} to quadratic order and diagonalizing the resulting mass matrix, one sees that, as asserted earlier, the theory has $N_F-1$ light modes and $N-N_F$ heavy modes. In particular, 
one sees that $\eta'$ is generically a heavy mode, and its mass is parametrically larger than the pseudo-NG bosons.  Setting the heavy modes in $\Sigma$ to their vacuum values, as well as taking $\eta'=0$, the $\theta$ dependent part of the effective potential \eqref{Master-3} reduce to standard chiral Lagrangian with its $\theta$ angle dependence: 
\begin{align}
V_{pNGB}
& = -  e^{-S_0}  f(m)  e^{i \theta/N_F } \tr \tilde U + c.c.
\label{NG}
\end{align}
The extrema (``vacuum family") of this potential   are  located at \footnote{These are the only configurations that are $SU(N_F)_V$ invariant. While we do not have $SU(N_F)_V$ symmetry, we are still able to focus on these configurations without loss of generality due to the fact that with our boundary conditions the Weyl group $S_N$ remains a symmetry.    }
\begin{align}
\tilde U_k= e^{i 2 \pi k/N_F}  {\bf 1}
\label{extremum}
\end{align}
 Indeed, 
using \eqref{extremum}, we find that the $\theta$ angle dependence of the vacuum energy as 
\begin{align}
E_{\rm vac} (\theta) = {\rm Min}_k E_{k} (\theta), \qquad 
E_{k} (\theta)
 = -  e^{-S_0}  f(m)  \cos \left( \frac{ \theta + 2 \pi k }{N_F} \right) 
\label{vacuum-energy}
\end{align}
For $\theta=0$, the vacuum is unique. For $\theta=\pi$,  $k=0$ and $k=-1$ are exactly degenerate, and CP is spontaneously broken for all $m_qe^{i \pi} <0$.  The point $m_q=0$ is a second order critical point in the phase diagram.

 \subsection{Special case with $N_F=2$}
 
The above discussion holds as written for $N \geq 3$ and  $N_F \geq 2$. However, the case $N_F=2$ requires special consideration, especially if one also sets $N=2$.  

Consider \eqref{Master} for $N =2 $,  $ N_F= 2$.  On $\mathbb{R}^4$ one expects five Nambu-Goldstone  bosons in the chiral limit due to the symmetry breaking pattern $SU(2N_F = 4) \to Sp(2N_F = 4)$, due to  the pseudo-reality of the fundamental representation for $N=2$.  But our twisted boundary conditions break this symmetry to its maximal Abelian subgroup, $SU(4) \to U(1)_L \times U(1)_R \times U(1)_Q$.  So at small $L$ we should still expect only one gapless mode at $m_q = 0$.  

Indeed, for $N =2$ there is only one dual photon, with the leading-order effective potential 
 \begin{align}
V(\sigma)= -  e^{-S_0}  m  \cos(\sigma) \cos (\theta/2 ) =  -  e^{-S_0}  f(m)  e^{i \theta/2} \tr U  + {\rm c.c.}  + \ldots
\label{2-2}
\end{align}
which vanishes at $\theta=\pi$. As we already discussed, at second order in semi-classical expansion, we must include the effects of magnetic bions, which induce a new term  in the effective potential: 
 \begin{align}
V(\sigma) & = -  e^{-S_0}  m \cos (\theta/2 )  \cos(\sigma) -e^{-2S_0}  |m|^2  \cos(2 \sigma)  + \ldots \cr \cr
& =  -  e^{-S_0}   m \cos( \theta/2) \tr U    -e^{-2S_0}  |m|^2   ( \tr U )^2 + \ldots
\label{2-2}
\end{align}
inducing a double-trace operator in the chiral Lagrangian. The chiral Lagrangian at second order semi-classics has  two vacua at 
$\theta=\pi$.  From the point of view of the chiral Lagrangian, these two vacua are stabilized by the double-trace term induced by the magnetic bions.
 
For $N_F =2, N \geq 3$,  \eqref{Master} and its equivalent re-writing \eqref{Master-3} includes both chiral Lagrangian modes and heavier modes.  There is also a contribution from magnetic bions, which can be interpreted as generating higher order terms in the chiral Lagrangian.  Minimizing the resulting potential, with second-order terms included, we again obtain  two-fold degenerate vacua at $\theta=\pi$. 

\subsection{$\theta$-dependence transmutation: How does $\frac{\theta}{N_F}$ change into $\frac{\theta}{N}$? }
\label{sec:transmutation}

One may wonder how the  $e^{i \theta/N}$  factor appearing in the  monopole operators $e^{-S_0}    e^{i \vec\alpha_i \cdot \vec\sigma }  e^{i \theta/N}$  in pure YM theory  gets  transmuted into the factors of $e^{i\theta/N_F}$ expected from a chiral Lagrangian relevant for small quark masses.    This question is hard to address on $\R^4$, but can be answered quite explicitly on $\R^3\times S^1$. 
                      
                     Consider the monopole operators in pure YM theory with the  inclusion of the $\theta$ angle. Since the topological charge of the monopole events is  $\frac{1}{N}$, they can be written as 
            \begin{align}
            e^{-S_0}    e^{i \vec\alpha_i \cdot \vec\sigma }  e^{i \theta/N},  \qquad i \in {\cal S}   
                     \end{align} 
       where               ${\cal S}$ is the affine root system of the Lie algebra. 
  If we introduce some massless fermions,  $\theta$ is redundant, as it can be eliminated from the path integral by a fermion field redefinition thanks to the chiral anomaly.   Without doing this field redefinition --- that is, keeping the ``fake" $\theta$ dependence ---  the monopole operators in the theory with $N_F$ fermions take the form: 
 \begin{align}
 &            e^{-S_0}  e^{i \vec\alpha_i \cdot \vec\sigma } 
       ({\psi}_{L,i} \psi_{R,i})  e^{i \theta/N},  \qquad  i \in  \, {\cal S}_1  \cr 
&       e^{-S_0}   e^{i \vec\alpha_k \cdot \vec\sigma } e^{i \theta/N}, 
                     \qquad \;\;\;\;\;\;\, \qquad   k  \in  \, {\cal S}_2 
 \end{align}  
Without loss of generality, assume that the monopole associated with the affine root lives in  ${\cal S}_1$, so that it has a fermion zero mode.  By globally shifting 
\begin{equation}
 {\vec \sigma} \rightarrow { \vec \sigma} - \frac{\theta}{N} 
 {\vec \rho} 
 \label{shift}
 \end{equation}
 where $ {\vec \rho}$ is the Weyl vector, we can remove the $\theta$ dependence from all of the operators appearing in  $ {\cal S}={\cal S}_1 \,\cup\,{\cal S}_2$, except for  $ {\cal M}_{N} $, for which $\theta$ dependence now looks like   
 \begin{align}
 e^{-S_0}  e^{i \vec\alpha_{N} \cdot \vec\sigma } 
       ({\psi}_{L,N} \psi_{R,N})  e^{i \theta},  \qquad N \in  \, {\cal S}_1  \, .
  \end{align} 
  Now, if we  globally shift  $  ({\psi}_{L,N} \psi_{R,N})  \rightarrow 
      e^{-i \theta}   ({\psi}_{L,N} \psi_{R,N})  $, the  $\theta$ dependence disappears from the monopole operators and the effective Lagrangian. Indeed, in the massless limit, this is the expected result.

      To see how $\theta/N_F$ in   \eqref{monopoles-3} arises, first do a field redefininition,  
      \begin{align}
 &       ({\psi}_{L,i} \psi_{R,i}) 
      \rightarrow    e^{i \theta/N_F}   ({\psi}_{L,i} \psi_{R,i}),   \quad \forall i \in {\cal S}_1 \setminus  {\cal M}_{N}  \cr 
    &   ({\psi}_{L,N} \psi_{R,N})  \rightarrow 
      e^{i(-1+ 1/N_F)  \theta }   ({\psi}_{L,N} \psi_{R,N})  
      \end{align}
Now, suppose we turn on a bare mass term in the 4D QFT, $m_q  \sum_{i \in {\cal S}_1} (\psi_{L,i} \psi_{R,i})$.  Then the fermion bilinear in the prefactor of the monopole operator should be replaced by the fermion mass.
As a result,  we obtain: 
\begin{align}
\label{monopoles-3}
{\cal M}_{i} &=    e^{-S_0}  e^{i \vec\alpha_i \cdot \vec\sigma } 
      m_q e^{i \theta/N_F}  , \qquad i \in  \, {\cal S}_1  ,  \cr
{\cal M}_{k} &=  e^{-S_0}   e^{i \vec\alpha_k \cdot \vec\sigma }   , 
                                  \qquad \qquad   \;\;\;\;\;\,  k  \in  \, {\cal S}_2  \, . 
\end{align}
starting from the monopole operators in YM theory. 

This transmutation is  the reason why observables depend on $\frac{\theta + 2 \pi k}{N_F} $ when the physics can be described by the softly broken chiral Lagrangian\cite{Rosenzweig:1979ay,DiVecchia:1980yfw,Witten:1980sp}, but come to depend on  $\frac{\theta + 2 \pi k}{N} $ when the quark mass is large enough that the physics is better described by pure Yang-Mills theory\cite{Witten:1978bc}. In other words, this explains why the functions describing observables have $N_F$ branches close to chiral limit, but come to have $N$ branches  away from that limit.

\section{From weak to strong coupling:  mixed anomalies and semiclassics } 
The analysis of the previous section shows that the properties of QCD-like gauge theories even  at arbitrarily small $\R^3 \times S^1$  are consistent with  the  expected properties of the same theories on $\R^4$.  This strongly supports the conjecture of adiabatic continuity of the physics as a function of $L$.   In this section, we present some strong  further evidence for the adiabatic continuity idea in $N_F$-flavor QCD. In particular,  we will show once the behavior of the theory is determined dynamically on small $\R^3 \times S^1$ by semi-classical methods, its behavior at arbitrarily large $S^1$, namely at strong coupling, is dictated by certain discrete 't Hooft anomalies.   

Before explaining the argument, we would like to emphasize that consideration of anomalies and the use of controlled semiclassical analysis are complementary approaches: 
  \begin{itemize}
 \item Studies of 't Hooft anomalies can determine the consistent choices of low-energy behavior of a QFT given some UV data.  But  typically there is more than one consistent options, and so one cannot determine the actual vacuum structure just from the anomalies, without making further assumptions about the dynamics. On the other hand, the constraints implied by the existence of 't Hooft anomalies are valid both at weak and strong coupling, which is very useful.
 
 \item Reliable semi-classics is a dynamical framework, and when it applies the low-energy behavior of a QFT can be conclusively determined.  However, this approach is only valid  at weak coupling. 
\end{itemize}
Therefore, the complementary nature of these two approaches should be clear, and it is natural to use them together.  Our arguments below will show that adiabatic continuity can be proved based on a single assumption in some of the theories of our interest.

In SU(N) QCD with $N_F$ Dirac flavors and the twisted boundary conditions explained around \eqref{eq:TBCs} there is an unbroken continuous chiral symmetry $[U(1)_A]^{N_F-1}$ in the chiral limit.  This continuous symmetry has an 't Hooft anomaly descending from the standard anomaly for $SU(N_F)_A$.  This anomaly implies that any long-distance effective field theory description of the theory must either include $N_F-1$ Nambu-Goldstone bosons, or it must contain a scale-invariant sector involving enough chiral fermions to completely saturate the anomaly.   A mixed long-distance phase, where the anomaly is partially satisfied by Nambu-Goldstone bosons and partly by chiral fermions, can be ruled out by the presence of the $S_{N_F}$ Weyl symmetry permuting the quark flavors, which cannot be spontaneously broken since it is a vector-like symmetry \cite{Vafa:1983tf}.

The results described in \cite{Cherman:2016hcd} and in the preceding sections of this paper are certainly consistent with these continuous 't Hooft anomalies.  Our small-$L$ results, as well as lattice results at large $L$, imply that the $[U(1)_A]^{N_F}$ symmetry is spontaneously broken.  However, a consideration of these continuous anomalies along with the large and small $L$ results does not directly rule out the exotic possibility that there is some window of values of $L$ where chiral symmetry is restored, and gapless chiral fermion excitations reemerge, before disappearing as $L$ is increased.  What does it take to exclude this exotic option for the dynamical behavior of QCD?

Here we argue that this option is severely constrained by consideration of \emph{discrete} 't Hooft anomalies.  Discrete 't Hooft anomalies in QCD and related theories have been the focus of several recent papers\cite{Gaiotto:2017yup,Komargodski:2017keh,Gaiotto:2017tne,Tanizaki:2017bam,Kikuchi:2017pcp,Tanizaki:2017qhf}.  Here we point out that there is a discrete 't Hooft anomaly that is of particular relevance to the behavior of CFC-symmetry-stabilized QCD as a function of $L$.  

The symmetries of QCD with $N_F$ massless Dirac fundamental fermions include a $\Z_{2N_F}$ flavor symmetry which is the discrete remnant of $U(1)_A$ axial symmetry.  These theories also possess a  color-flavor center symmetry, $\mathbb{Z}_d$ provided 
$d = \mathrm{gcd}(N,N_F) >1$.  There is a mixed 't Hooft anomaly between these two symmetries. To see this, we first recall why $\Z_{2N_F}$  axial symmetry is respected.  For example, in a color gauge field background, the equation describing
axial charge non-conservation can be written as 
 \begin{align}
 \Delta Q_5= 2N_F  \times  {1 \over 8\pi^2} \int 
\tr( F\wedge F)  \in 2N_F  \Z 
\end{align}
The second relation follows from the quantization of the topological charge, implying that  $\Z_{2N_F}$ is an exact symmetry of the theory.  

In order to show the existence of the claimed mixed 't Hooft anomaly, suppose we gauge the $\mathbb{Z}_d$ CFC-symmetry.  To describe how this can be done, let us consider the simplest example, $N_F=N$.  Then let us first suppose that we gauge the flavor symmetry $SU(N_F)_V$.  The result can be thought of as a two-site quiver gauge theory.  We can then gauge the $\mathbb{Z}_N$ global center symmetry.  (For general $N_F$, this would be a $\mathbb{Z}_{d}$ symmetry.)  To do this, one can replace the two $SU(N)$ connections $F_1, F_2$ by $U(N)$ connections $F'_1,F'_2$ while at the same time adding some Lagrange multiplier gauge fields to the path integral to restrict the gauge transformations to the desired discrete subgroup.  This procedure is described in e.g. \cite{Gaiotto:2017tne,Gaiotto:2017yup,Tanizaki:2017bam}.  In the end, the relation between the sum of the topological $\theta$ terms and the axial charge non-conservation becomes
 \begin{align}
&\frac{1}{ 8\pi^2} \sum_{i = 1,2} \int 
\left( \tr F'_i \wedge F'_i -  \frac{1}{ N } \tr F'_i \wedge \tr F'_i   \right)  \in \frac{2}{N} \Z \,. 
\end{align}
Consequently, 
\begin{align}
\Delta Q_5  \in   4 \Z
\end{align}
and one loses most of the discrete chiral symmetry upon gauging the CFC symmetry.  Therefore,  there  exists a mixed anomaly between the $\mathbb{Z}_d$ CFC symmetry and the $\mathbb{Z}_{2N_F}$ symmetry.

This mixed 't Hooft anomaly implies that either  $\mathbb{Z}_d$ or $\Z_{2N_F}$ (or both) are spontaneously broken, or  the long-distance EFT includes some appropriate TQFT.    But  one cannot have a phase in which  the ground state is unique (``trivial") on arbitrary manifolds.   This is sometimes referred to as persistent order. The concept of adiabatic continuity is a refined  version of the persistent order where only one mode of the symmetry realization holds throughout a range of parameters. 

In the small circle regime, we have ensured that the $\mathbb{Z}_d$ CFC symmetry is unbroken. Therefore, the  only anomaly-consistent dynamics must drive spontaneous breaking of the
$\mathbb{Z}_{2N_F}$ axial symmetry. This is indeed the case, as we show momentarily. However, the full story is more interesting than that.
The interesting aspect is following:

\begin{itemize}
\item Although here the relevant 't Hooft anomaly involves two \emph{discrete} symmetries,  there are no order parameters that  
transform under  $\mathbb{Z}_{2N_F}$, but not under maximal abelian chiral symmetry $(U(1)_A)^{N_F-1}$. Therefore, the discrete 't Hooft anomaly actually means that an unbroken 
$\mathbb{Z}_d$ CFC symmetry implies the {\it complete} spontaneous breaking of continuous chiral symmetry. 
\end{itemize}

Indeed, our earlier semi-classical analysis on small  $\R^3 \times S^1$ shows that in the regime where $\mathbb{Z}_d$ is unbroken, 
maximal abelian chiral symmetry $[U(1)_A]^{N_F-1}$ is broken. In this regime, we derived the chiral Lagrangian \eqref{Master-3}
from microscopic dynamics. Now, {\it with the assumption} that  $\mathbb{Z}_d$  is unbroken at any circle size,  $[U(1)_A]^{N_F-1}$ must be spontaneously broken at any radius, including at large radius where the theory is strongly coupled!  The assumption of an unbroken CFC symmetry can be guaranteed for any circle size by adding a CFC-stabilizing double-trace deformation to the theory.   We believe this provides a derivation of the chiral Lagrangian at both weak and strong coupling in the class of theories under consideration. 

Finally, we note that the value of doing controlled semiclassical calculations and systematic effective field theory analysis is that by construction, such methods cannot produce results that are inconsistent with the structure of a given well-defined quantum field theory.  This should be contrasted with ad hoc models based on uncontrolled approximations, such as e.g. four-dimensional NJL models, which are far less constrained, and can easily produce incorrect results.

\section{Three dimensions versus four dimensions}
Our discussion above focused on analyzing locally four-dimensional $SU(N)$ gauge theories with $N_F$ fundamental quarks with a common mass $m_q$.  In a semiclassically-calculable regime on $\mathbb{R}^3 \times S^1$, where the $S^1$ size $L$ is small, we saw that these theories become gapless at \emph{points} in parameter space.    In particular, when $N_F = 1$, we saw that there is a single critical point at $m_q/\Lambda = m^{*} \in \mathbb{R}^{-}$, while for $N_F >1$, the critical point is at $m^{*} = 0$.

To highlight the critical role that the four-dimensional nature of these theories played in our results above, as well as for its independent theoretical interest, we now consider a three-dimensional $SU(2)$ gauge theory with $N_F = 2$ Dirac\footnote{That is, complex two-component 3D spinors.} fundamental quarks $\psi_{a}, a=1,2$, as well as an adjoint  scalar $\phi$.  Roughly speaking, this  theory arises as a naive dimensional reduction of the 4d   $N_F = 1$ two-color theory down to 3d. 

 We choose the scalar potential so that the $SU(2)$ gauge symmetry is spontaneously `broken' to $U(1)$ at long distances via the adjoint Higgs mechanism.  So we take the Euclidean action of the theory to be 
\begin{align}
\label{eq:3dQCD}
S_3 = \int d^3{x} \,&\bigg[ \frac{1}{2g^2}\tr F_{\mu\nu}F^{\mu\nu} +  \tr D_{\mu} \phi D^{\mu} \phi  + \lambda \tr (\phi^2 - v^2)^2 + \\
&  \sum_{a=1,2} \left(\bar{\psi}_a  \Dslash \psi_a \right) +  m_q (\bar{\psi}_1\psi_1- \bar{\psi}_2 \psi_2) \bigg] \, .
\end{align}
where $D_{\mu} \phi = \partial_{\mu}\phi + i [A_{\mu},\phi], D_{\mu} \psi = \partial_{\mu}\psi_a + i A_{\mu}\psi_a$, $A_{\mu}$ is the $SU(2)$ gauge field, $F_{\mu \nu}$ is its field strength, and $v^2 >0$. With the given Higgs potential, the $SU(2)$ gauge symmetry is ``spontaneously broken" to $U(1)$, with a W-boson mass scale $m_W = \frac{1}{2} g v$.  In what follows we assume that  $m_q \gtrsim g^2$, and at the same time also take $v \gg g$.  Then the long-distance theory is \emph{weakly-coupled} 3D compact QED coupled to two flavors of  massive `electrons' with opposite charges, meaning that the gauge symmetry acts as 
\begin{align}
U(1)_{\rm gauge}\, : \qquad\psi \to e^{i\alpha(x) \tau_3} \psi \,.
\end{align}
where $\psi$ is a flavor doublet and $\tau_3 = \mathrm{diag}(1,-1)$ is a flavor matrix.  Note that the masses of the $\psi_a$ fields can easily be smaller than the UV cutoff scale $m_W$ without any loss of control over the long-distance dynamics, given the assumed hierarchies for the parameters of \eqref{eq:3dQCD}. \footnote{There is also a Higgs mode with mass $m_H = \lambda v^2$.  We assume that $m_H \gtrsim m_W$ and  ignore it in our discussion of the low-energy behavior. }   
We should also emphasize that with the mass term we have chosen, the theory is parity-invariant.  Indeed, the two fermion flavors give opposite-sign contributions to the 3D parity anomaly\cite{Redlich:1983dv}.  Therefore the parity anomaly cancels.  

Because of the absence of the chiral anomaly in 3d,  the anomalous $\Z_{2N_F}$ discrete chiral symmetry of the analogous 4D theory enhances 
to a $U(1)$ global symmetry in our three dimensional theory.  So the continuous global symmetry of our theory is essentially
\begin{align} 
&U(1) :   \qquad  \;\;  \psi_a \rightarrow e^{i \alpha} \psi_a \,.
\label{sym}
\end{align}
More precisely, the global symmetry is $U(1)/\mathbb{Z}_2$, where  the $\mathbb{Z}_2$ quotient acts by $\psi_a \to -\psi_a$.  Note that even though this $U(1)$ symmetry can be viewed as an ``axial" symmetry in the context of a 4d uplift of the 3d theory, from an intrinsically 3d perspective the $U(1)$ symmetry is actually the quark number symmetry, in 3d QCD language.

\begin{figure}
\centering
\includegraphics[width=0.8\textwidth]{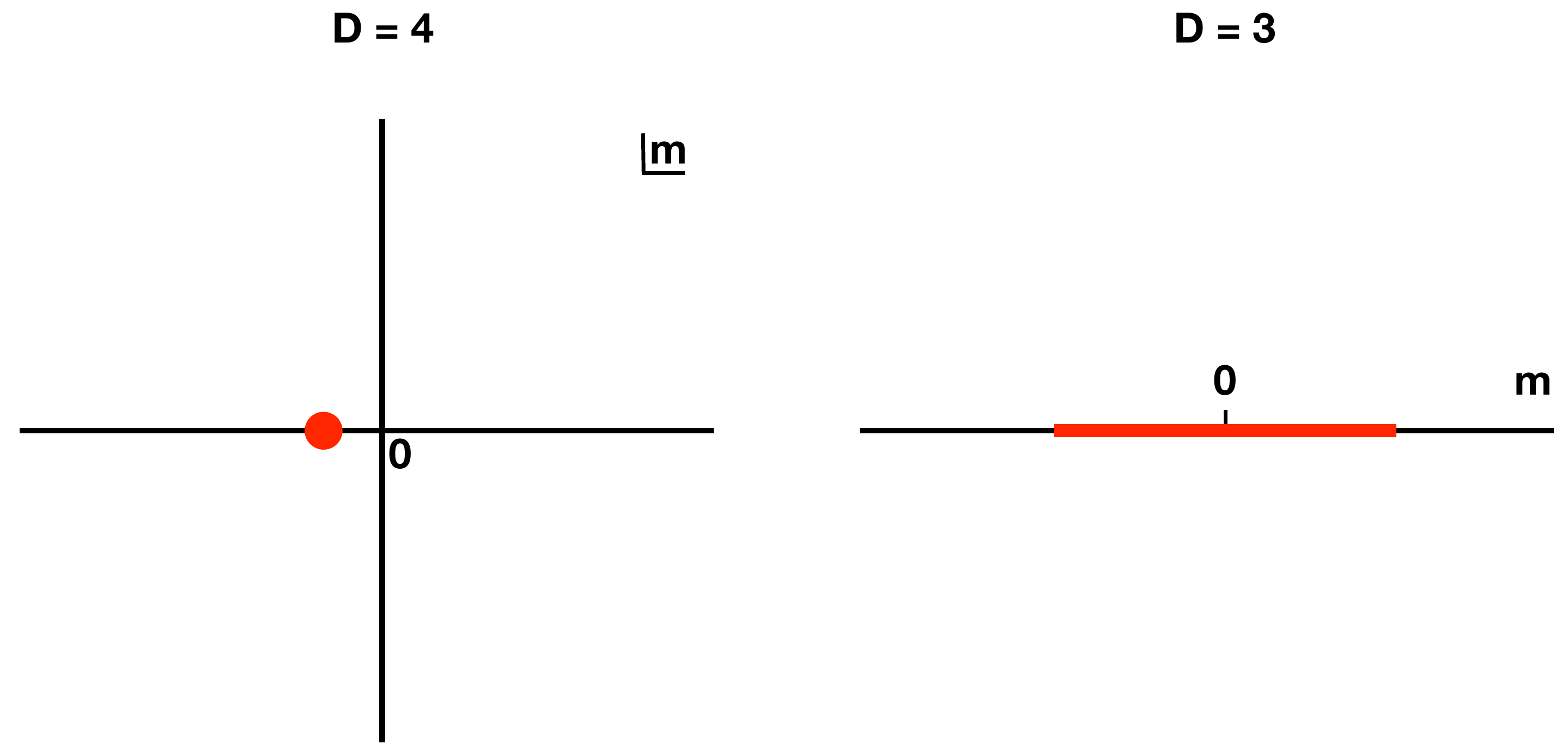}
\caption{In four-dimensional QCD-like theories, we find that the critical region in mass-parameter space is a point, which is illustrated on the left for the $N_F=1$ theory.  On the other hand, in three-dimensional theories, the critical region can be an interval as a function of a parity-invariant mass term, as is shown explicitly in the main text in a simple example.
}
\label{fig:fm_sketch}
\end{figure}

If not for the presence of fundamental fermions, we would be dealing with the theory analyzed by Polyakov in his classic paper\cite{Polyakov:1976fu}.  Taking into account the contributions of monopole-instantons, Polyakov found that this theory has a non-perturbative mass gap $\Delta \sim m_W e^{-4\pi m_W/g^2}$. Note that $\Delta$ is parametrically smaller than the cutoff scale $m_W$ of the 3D QED effective field theory.   As a consequence, the low-energy behavior physics of compact QED is completely different from that of non-compact QED.  Non-compact 3d QED without matter has no monopole-instantons, and hence it is a free gapless theory in the absence of matter.  But compact 3d QED without matter is always gapped.  The coupling of Polyakov's theory to \emph{adjoint} fermions was analyzed by Affleck, Harvey, and Witten\cite{Affleck:1982as}, with the result that the theory is gapless at a point in parameter space, which is $m_q = 0$.\footnote{In locally four-dimensional gauge theories with adjoint fermions the situation is quite different. There massless adjoint fermions do not necessarily render the long-distance behavior gapless, due to the magnetic bion mechanism\cite{Unsal:2007jx}.}   

 We now show that the behavior of \eqref{eq:3dQCD}, which contains fundamental fermions, is completely different, and instead of a critical point in parameter space, there is instead a \emph{critical interval}, given by 
\begin{align}
m_q \in \mathcal{C} = \left(-g v,+g v \right) \,.
\label{eq:gapless_interval}
\end{align}  
In this interval, the theory is gapless.  More precisely, below we show that this is true so long as $|m| \gtrsim g^2$, where the theory is weakly coupled.  It is logically possible that a gap appears in some intervals within the strongly-coupled region $|m| \lesssim g^2$.  If so, the theory would have multiple critical intervals.  In any case, everywhere in $\mathcal{C}$ where we can trust our calculations, the $U(1)$ quark number symmetry is spontaneously broken, and the gapless mode is in fact a Nambu-Goldstone boson.  If  $m_q$ is outside of the range \eqref{eq:gapless_interval}, then the theory has a gap, and the $U(1)$ symmetry is not broken. Since in 3D the dual photon is parity even, parity is not spontaneously broken anywhere on the $m_q \in \mathbb{R}$ line as long as we can trust our semiclassical analysis.

We note that this result is consistent with the conjectured scenario of \cite{Komargodski:2017smk} in some closely related theories.  In the present setting, the existence of a critical interval in parameter space can actually be shown explicitly, rather than being conjectured.  To see this, one can use the Callias index theorem.  Unlike in the four-dimensional $SU(2)$ gauge theory considered in a previous section, where there are two different monopole events thanks to the existence of the KK monopole-instanton, when the adjoint Higgs field is non-compact and algebra-valued, as is the case here, there is only \emph{one} monopole event. 
The Callias index theorem \cite{Callias:1977kg} says that the number of fermionic zero modes  on the monopole background depends on the relation between $m_q$ and  adjoint Higgs vacuum expectation value, and is given by 
\begin{align}
I_{\rm Callias} = \left\{   \begin{array}{l l}
 2 \qquad &  |m_q| < gv    \cr
0 \qquad &    |m_q| > g v
\end{array} \right.
\end{align}
For an explicit demonstration of related behavior in the 4D context from solutions to the Dirac equation in a monopole background, see e.g. \cite{Csaki:2017cqm}. The jump in the 3D index as the boundary in parameter space is crossed has important implications in 3d dynamics.

The jump in the index implies that the form of the monopole operator changes as the boundary $|m_q|=gv$ is crossed.  
\begin{align}
\cal M = \left\{   \begin{array}{l l}
e^{-S_0} e^{i \sigma} \psi_1 \psi_2  \qquad &  |m_q| < gv     \cr
e^{-S_0} e^{i \sigma}   \qquad &    |m_q| > g v
\end{array} \right.
\end{align}
For $m_q \in \mathcal{C}$, the $U(1)$  symmetry  (which is non-anomalous in 3d) intertwines with the shift symmetry of the dual photon, $U(1)_J$.   Under  the diagonal of $U(1)$ and $U(1)_J$,  one has
 \begin{align} 
&[U(1)]_{\rm diag} :   \qquad  \;\;  \psi_1 \psi_2  \rightarrow e^{i 2 \alpha}   \psi_1 \psi_2, \qquad   \sigma  \rightarrow \sigma -2 \alpha 
\label{sym-2}
\end{align}
So the dual photon has an exact shift symmetry.  This shift symmetry is spontaneously broken by any vacuum expectation value for $\sigma$.  Thus we find that in the interval \eqref{eq:gapless_interval}, the theory is gapless and has a spontaneously-broken $U(1)$ quark-number symmetry. The dual photon is the Nambu-Goldstone boson.
Outside the interval  $\mathcal{C}$, $U(1)_J$ is explicitly broken by the monopole induced potential  $V(\sigma)= - e^{-S_0} \cos\sigma $, while  the $U(1)$ quark number symmetry is not spontaneously broken.
These remarks imply that there are at least two phase transitions as the mass parameter is dialed from positive to negative, in such a way that there is at least one critical interval in $m_q$, as we claimed at the beginning of this section.

\subsection{Implications for 3D compact QED}
The study of the long-distance properties of \emph{non}-compact QED has recently been revitalized by advances in $\epsilon$ expansions and modern lattice simulations, see e.g. \cite{DiPietro:2015taa,Giombi:2015haa,Karthik:2015sgq,Karthik:2016ppr,DiPietro:2017kcd}.  In these papers, all of the $n_F$ fermion flavors are assigned the same charge under the gauge symmetry. Consequently, one deals with theories with an $SU(n_F)$ global symmetry. The lattice calculations indicate that in the non-compact version of these theories there is no sign of spontaneous symmetry breaking for  $SU(n_F)$.

However, in the condensed-matter applications of 3d QED,  typically it is the \emph{compact} version of 3d QED that arises\cite{Marston:1989zz,Rantner:2000wer,Rantner:2002zz,Hermele:2005dkq,Hermele:2008afa,Ran:2007zea}.  Our analysis above implies that in parity-invariant compact 3d QED with two flavors of electrons with opposite charges, the global $U(1)$ symmetry is spontaneously broken in an interval in the parameter space of the model.  It would be interesting to work out the implications of this observation in a condensed-matter context.  It would also be interesting to understand the effects of also turning on a parity-breaking mass term, and subsequently to make closer contact with the analysis of \cite{Komargodski:2017keh}.

\acknowledgments
We are grateful to K.~Aitken,  Z.~Komargodski, T.~Sulejmanpasic, and L.~G.~Yaffe for helpful discussions.  We also thank the KITP for its warm hospitality as part of the program ``Resurgent Asymptotics in Physics and Mathematics", where this paper was completed.  Research at KITP is supported by the National Science Foundation under Grant No. NSF PHY11-25915.  A.~C. was also supported by the U. S. Department of Energy via grants DE-FG02-00ER-41132 (A.C.), while M.~\"U. was supported U. S. Department of Energy grant  DE-FG02-03ER41260.

As the write-up of this paper was being finalized, we became aware of a forthcoming paper\cite{TKMS_unpublished} analyzing finite quark density physics of QCD in the context of a $\mathbb{Z}_{N_F}$-twisted partition function, which has some overlap with some of our discussion.  We thank Y.~Kikuchi, T.~Misumi, N.~Sakai, and Y.~Tanizaki for sharing a draft of their paper prior to publication, and for helpful comments on a draft of our paper.

\appendix


\bibliography{small_circle}

\providecommand{\href}[2]{#2}\begingroup\raggedright\begin{thebibliography}{10}

\bibitem{Gaiotto:2017tne}
D.~Gaiotto, Z.~Komargodski and N.~Seiberg, \emph{{Time-Reversal Breaking in
  QCD$_4$, Walls, and Dualities in 2+1 Dimensions}},
  \href{https://arxiv.org/abs/1708.06806}{{\ttfamily 1708.06806}}.

\bibitem{Polyakov:1976fu}
A.~M. Polyakov, \emph{{Quark Confinement and Topology of Gauge Groups}},
  \href{https://doi.org/10.1016/0550-3213(77)90086-4}{\emph{Nucl. Phys.}
  {\bfseries B120} (1977) 429--458}.

\bibitem{0022-3719-10-19-011}
J.~M. Kosterlitz, \emph{The d-dimensional coulomb gas and the roughening
  transition}, {\emph{Journal of Physics C: Solid State Physics} {\bfseries 10}
  (1977) 3753}.

\bibitem{Read:1990zza}
N.~Read and S.~Sachdev, \emph{{Spin-Peierls, valence-bond solid, and Neel
  ground states of low-dimensional quantum antiferromagnets}},
  \href{https://doi.org/10.1103/PhysRevB.42.4568}{\emph{Phys. Rev.} {\bfseries
  B42} (1990) 4568--4589}.

\bibitem{Fradkin:1991nr}
E.~H. Fradkin, \emph{{Field Theories of Condensed Matter Physics}},
  {\emph{Front. Phys.} {\bfseries 82} (2013) 1--852}.

\bibitem{Cherman:2016hcd}
A.~Cherman, T.~Sch{\"a}fer and M.~{\"U}nsal, \emph{{Chiral Lagrangian from
  Duality and Monopole Operators in Compactified QCD}},
  \href{https://doi.org/10.1103/PhysRevLett.117.081601}{\emph{Phys. Rev. Lett.}
  {\bfseries 117} (2016) 081601},
  [\href{https://arxiv.org/abs/1604.06108}{{\ttfamily 1604.06108}}].

\bibitem{Cherman:2017tey}
A.~Cherman, S.~Sen, M.~Unsal, M.~L. Wagman and L.~G. Yaffe, \emph{{Order
  parameters and color-flavor center symmetry in QCD}},
  \href{https://arxiv.org/abs/1706.05385}{{\ttfamily 1706.05385}}.

\bibitem{Gaiotto:2017yup}
D.~Gaiotto, A.~Kapustin, Z.~Komargodski and N.~Seiberg, \emph{{Theta, Time
  Reversal, and Temperature}},
  \href{https://doi.org/10.1007/JHEP05(2017)091}{\emph{JHEP} {\bfseries 05}
  (2017) 091}, [\href{https://arxiv.org/abs/1703.00501}{{\ttfamily
  1703.00501}}].

\bibitem{Unsal:2007vu}
M.~Unsal, \emph{{Abelian duality, confinement, and chiral symmetry breaking in
  QCD(adj)}},
  \href{https://doi.org/10.1103/PhysRevLett.100.032005}{\emph{Phys.Rev.Lett.}
  {\bfseries 100} (2008) 032005},
  [\href{https://arxiv.org/abs/0708.1772}{{\ttfamily 0708.1772}}].

\bibitem{Unsal:2007jx}
M.~Unsal, \emph{{Magnetic bion condensation: A New mechanism of confinement and
  mass gap in four dimensions}},
  \href{https://doi.org/10.1103/PhysRevD.80.065001}{\emph{Phys. Rev.}
  {\bfseries D80} (2009) 065001},
  [\href{https://arxiv.org/abs/0709.3269}{{\ttfamily 0709.3269}}].

\bibitem{Unsal:2008ch}
M.~Unsal and L.~G. Yaffe, \emph{{Center-stabilized Yang-Mills theory:
  Confinement and large N volume independence}},
  \href{https://doi.org/10.1103/PhysRevD.78.065035}{\emph{Phys. Rev.}
  {\bfseries D78} (2008) 065035},
  [\href{https://arxiv.org/abs/0803.0344}{{\ttfamily 0803.0344}}].

\bibitem{Shifman:2008ja}
M.~Shifman and M.~Unsal, \emph{{QCD-like Theories on R(3) x S(1): A Smooth
  Journey from Small to Large r(S(1)) with Double-Trace Deformations}},
  \href{https://doi.org/10.1103/PhysRevD.78.065004}{\emph{Phys. Rev.}
  {\bfseries D78} (2008) 065004},
  [\href{https://arxiv.org/abs/0802.1232}{{\ttfamily 0802.1232}}].

\bibitem{Shifman:2009tp}
M.~Shifman and M.~Unsal, \emph{{Multiflavor QCD* on R(3) x S(1): Studying
  Transition From Abelian to Non-Abelian Confinement}},
  \href{https://doi.org/10.1016/j.physletb.2009.10.060}{\emph{Phys. Lett.}
  {\bfseries B681} (2009) 491--494},
  [\href{https://arxiv.org/abs/0901.3743}{{\ttfamily 0901.3743}}].

\bibitem{Unsal:2010qh}
M.~Unsal and L.~G. Yaffe, \emph{{Large-N volume independence in conformal and
  confining gauge theories}},
  \href{https://doi.org/10.1007/JHEP08(2010)030}{\emph{JHEP} {\bfseries 08}
  (2010) 030}, [\href{https://arxiv.org/abs/1006.2101}{{\ttfamily 1006.2101}}].

\bibitem{Shifman:2008cx}
M.~Shifman and M.~Unsal, \emph{{On Yang-Mills Theories with Chiral Matter at
  Strong Coupling}},
  \href{https://doi.org/10.1103/PhysRevD.79.105010}{\emph{Phys. Rev.}
  {\bfseries D79} (2009) 105010},
  [\href{https://arxiv.org/abs/0808.2485}{{\ttfamily 0808.2485}}].

\bibitem{Cossu:2009sq}
G.~Cossu and M.~D'Elia, \emph{{Finite size phase transitions in QCD with
  adjoint fermions}},
  \href{https://doi.org/10.1088/1126-6708/2009/07/048}{\emph{JHEP} {\bfseries
  07} (2009) 048}, [\href{https://arxiv.org/abs/0904.1353}{{\ttfamily
  0904.1353}}].

\bibitem{Myers:2009df}
J.~C. Myers and M.~C. Ogilvie, \emph{{Phase diagrams of SU(N) gauge theories
  with fermions in various representations}},
  \href{https://doi.org/10.1088/1126-6708/2009/07/095}{\emph{JHEP} {\bfseries
  07} (2009) 095}, [\href{https://arxiv.org/abs/0903.4638}{{\ttfamily
  0903.4638}}].

\bibitem{Simic:2010sv}
D.~Simic and M.~Unsal, \emph{{Deconfinement in Yang-Mills theory through
  toroidal compactification with deformation}},
  \href{https://doi.org/10.1103/PhysRevD.85.105027}{\emph{Phys. Rev.}
  {\bfseries D85} (2012) 105027},
  [\href{https://arxiv.org/abs/1010.5515}{{\ttfamily 1010.5515}}].

\bibitem{Vairinhos:2011gv}
H.~Vairinhos, \emph{{Phase transitions in center-stabilized lattice gauge
  theories}}, {\emph{PoS} {\bfseries LATTICE2011} (2011) 252},
  [\href{https://arxiv.org/abs/1111.0303}{{\ttfamily 1111.0303}}].

\bibitem{Thomas:2011ee}
E.~Thomas and A.~R. Zhitnitsky, \emph{{Topological Susceptibility and Contact
  Term in QCD. A Toy Model}},
  \href{https://doi.org/10.1103/PhysRevD.85.044039}{\emph{Phys. Rev.}
  {\bfseries D85} (2012) 044039},
  [\href{https://arxiv.org/abs/1109.2608}{{\ttfamily 1109.2608}}].

\bibitem{Anber:2011gn}
M.~M. Anber, E.~Poppitz and M.~Unsal, \emph{{2d affine XY-spin model/4d gauge
  theory duality and deconfinement}},
  \href{https://doi.org/10.1007/JHEP04(2012)040}{\emph{JHEP} {\bfseries 04}
  (2012) 040}, [\href{https://arxiv.org/abs/1112.6389}{{\ttfamily 1112.6389}}].

\bibitem{Poppitz:2012sw}
E.~Poppitz, T.~Sch{\"a}fer and M.~Unsal, \emph{{Continuity, Deconfinement, and
  (Super) Yang-Mills Theory}},
  \href{https://doi.org/10.1007/JHEP10(2012)115}{\emph{JHEP} {\bfseries 10}
  (2012) 115}, [\href{https://arxiv.org/abs/1205.0290}{{\ttfamily 1205.0290}}].

\bibitem{Poppitz:2012nz}
E.~Poppitz, T.~Sch{\"a}fer and M.~{\"U}nsal, \emph{{Universal mechanism of
  (semi-classical) deconfinement and theta-dependence for all simple groups}},
  \href{https://doi.org/10.1007/JHEP03(2013)087}{\emph{JHEP} {\bfseries 03}
  (2013) 087}, [\href{https://arxiv.org/abs/1212.1238}{{\ttfamily 1212.1238}}].

\bibitem{Unsal:2012zj}
M.~Unsal, \emph{{Theta dependence, sign problems and topological
  interference}}, \href{https://doi.org/10.1103/PhysRevD.86.105012}{\emph{Phys.
  Rev.} {\bfseries D86} (2012) 105012},
  [\href{https://arxiv.org/abs/1201.6426}{{\ttfamily 1201.6426}}].

\bibitem{Argyres:2012ka}
P.~C. Argyres and M.~Unsal, \emph{{The semi-classical expansion and resurgence
  in gauge theories: new perturbative, instanton, bion, and renormalon
  effects}}, \href{https://doi.org/10.1007/JHEP08(2012)063}{\emph{JHEP}
  {\bfseries 08} (2012) 063},
  [\href{https://arxiv.org/abs/1206.1890}{{\ttfamily 1206.1890}}].

\bibitem{Argyres:2012vv}
P.~Argyres and M.~Unsal, \emph{{A semiclassical realization of infrared
  renormalons}},
  \href{https://doi.org/10.1103/PhysRevLett.109.121601}{\emph{Phys. Rev. Lett.}
  {\bfseries 109} (2012) 121601},
  [\href{https://arxiv.org/abs/1204.1661}{{\ttfamily 1204.1661}}].

\bibitem{Anber:2013doa}
M.~M. Anber, S.~Collier, E.~Poppitz, S.~Strimas-Mackey and B.~Teeple,
  \emph{{Deconfinement in $\mathcal{N}=1$ super Yang-Mills theory on
  $\mathbb{R}^3 \times \mathbb{S}^1$ via dual-Coulomb gas and "affine"
  XY-model}}, \href{https://doi.org/10.1007/JHEP11(2013)142}{\emph{JHEP}
  {\bfseries 11} (2013) 142},
  [\href{https://arxiv.org/abs/1310.3522}{{\ttfamily 1310.3522}}].

\bibitem{Cossu:2013ora}
G.~Cossu, H.~Hatanaka, Y.~Hosotani and J.-I. Noaki, \emph{{Polyakov loops and
  the Hosotani mechanism on the lattice}},
  \href{https://doi.org/10.1103/PhysRevD.89.094509}{\emph{Phys. Rev.}
  {\bfseries D89} (2014) 094509},
  [\href{https://arxiv.org/abs/1309.4198}{{\ttfamily 1309.4198}}].

\bibitem{Bhoonah:2014gpa}
A.~Bhoonah, E.~Thomas and A.~R. Zhitnitsky, \emph{{Metastable vacuum decay and
  $\theta$ dependence in gauge theory. Deformed QCD as a toy model}},
  \href{https://doi.org/10.1016/j.nuclphysb.2014.11.007}{\emph{Nucl. Phys.}
  {\bfseries B890} (2014) 30--47},
  [\href{https://arxiv.org/abs/1407.5121}{{\ttfamily 1407.5121}}].

\bibitem{Anber:2014lba}
M.~M. Anber, E.~Poppitz and B.~Teeple, \emph{{Deconfinement and continuity
  between thermal and (super) Yang-Mills theory for all gauge groups}},
  \href{https://doi.org/10.1007/JHEP09(2014)040}{\emph{JHEP} {\bfseries 09}
  (2014) 040}, [\href{https://arxiv.org/abs/1406.1199}{{\ttfamily 1406.1199}}].

\bibitem{Bergner:2014dua}
G.~Bergner and S.~Piemonte, \emph{{Compactified $ \mathcal{N}=1 $
  supersymmetric Yang-Mills theory on the lattice: continuity and the
  disappearance of the deconfinement transition}},
  \href{https://doi.org/10.1007/JHEP12(2014)133}{\emph{JHEP} {\bfseries 12}
  (2014) 133}, [\href{https://arxiv.org/abs/1410.3668}{{\ttfamily 1410.3668}}].

\bibitem{Li:2014lza}
X.~Li and M.~B. Voloshin, \emph{{Metastable vacuum decay in center-stabilized
  Yang-Mills theory at large N}},
  \href{https://doi.org/10.1103/PhysRevD.90.105028}{\emph{Phys. Rev.}
  {\bfseries D90} (2014) 105028},
  [\href{https://arxiv.org/abs/1408.3054}{{\ttfamily 1408.3054}}].

\bibitem{Anber:2015kea}
M.~M. Anber, E.~Poppitz and T.~Sulejmanpasic, \emph{{Strings from domain walls
  in supersymmetric Yang-Mills theory and adjoint QCD}},
  \href{https://doi.org/10.1103/PhysRevD.92.021701}{\emph{Phys. Rev.}
  {\bfseries D92} (2015) 021701},
  [\href{https://arxiv.org/abs/1501.06773}{{\ttfamily 1501.06773}}].

\bibitem{Anber:2015wha}
M.~M. Anber and E.~Poppitz, \emph{{On the global structure of deformed
  Yang-Mills theory and QCD(adj) on $ {\mathrm{\mathbb{R}}}^3\times
  {\mathbb{S}}^1 $}},
  \href{https://doi.org/10.1007/JHEP10(2015)051}{\emph{JHEP} {\bfseries 10}
  (2015) 051}, [\href{https://arxiv.org/abs/1508.00910}{{\ttfamily
  1508.00910}}].

\bibitem{Misumi:2014raa}
T.~Misumi and T.~Kanazawa, \emph{{Adjoint QCD on $\mathbb{R}^3\times S^1$ with
  twisted fermionic boundary conditions}},
  \href{https://doi.org/10.1007/JHEP06(2014)181}{\emph{JHEP} {\bfseries 06}
  (2014) 181}, [\href{https://arxiv.org/abs/1405.3113}{{\ttfamily 1405.3113}}].

\bibitem{Aitken:2017ayq}
K.~Aitken, A.~Cherman, E.~Poppitz and L.~G. Yaffe, \emph{{QCD on a small
  circle}},  \href{https://arxiv.org/abs/1707.08971}{{\ttfamily 1707.08971}}.

\bibitem{Anber:2017rch}
M.~M. Anber and A.~R. Zhitnitsky, \emph{{Oblique Confinement at $\theta\neq 0$
  in weakly coupled gauge theories with deformations}},
  \href{https://doi.org/10.1103/PhysRevD.96.074022}{\emph{Phys. Rev.}
  {\bfseries D96} (2017) 074022},
  [\href{https://arxiv.org/abs/1708.07520}{{\ttfamily 1708.07520}}].

\bibitem{Iritani:2015ara}
T.~Iritani, E.~Itou and T.~Misumi, \emph{{Lattice study on QCD-like theory with
  exact center symmetry}},
  \href{https://doi.org/10.1007/JHEP11(2015)159}{\emph{JHEP} {\bfseries 11}
  (2015) 159}, [\href{https://arxiv.org/abs/1508.07132}{{\ttfamily
  1508.07132}}].

\bibitem{Komargodski:2017keh}
Z.~Komargodski and N.~Seiberg, \emph{{A Symmetry Breaking Scenario for
  QCD$_3$}},  \href{https://arxiv.org/abs/1706.08755}{{\ttfamily 1706.08755}}.

\bibitem{Lee:1997vp}
K.-M. Lee and P.~Yi, \emph{{Monopoles and instantons on partially compactified
  D-branes}}, \href{https://doi.org/10.1103/PhysRevD.56.3711}{\emph{Phys. Rev.}
  {\bfseries D56} (1997) 3711--3717},
  [\href{https://arxiv.org/abs/hep-th/9702107}{{\ttfamily hep-th/9702107}}].

\bibitem{Kraan:1998sn}
T.~C. Kraan and P.~van Baal, \emph{{Monopole constituents inside SU(n)
  calorons}}, \href{https://doi.org/10.1016/S0370-2693(98)00799-0}{\emph{Phys.
  Lett.} {\bfseries B435} (1998) 389--395},
  [\href{https://arxiv.org/abs/hep-th/9806034}{{\ttfamily hep-th/9806034}}].

\bibitem{Dunne:2004sx}
G.~V. Dunne, J.~Hur, C.~Lee and H.~Min, \emph{{Precise quark mass dependence of
  instanton determinant}},
  \href{https://doi.org/10.1103/PhysRevLett.94.072001}{\emph{Phys. Rev. Lett.}
  {\bfseries 94} (2005) 072001},
  [\href{https://arxiv.org/abs/hep-th/0410190}{{\ttfamily hep-th/0410190}}].

\bibitem{Affleck:1982as}
I.~Affleck, J.~A. Harvey and E.~Witten, \emph{{Instantons and (Super)Symmetry
  Breaking in (2+1)-Dimensions}},
  \href{https://doi.org/10.1016/0550-3213(82)90277-2}{\emph{Nucl. Phys.}
  {\bfseries B206} (1982) 413--439}.

\bibitem{Kouno:2012zz}
H.~Kouno, Y.~Sakai, T.~Makiyama, K.~Tokunaga, T.~Sasaki and M.~Yahiro,
  \emph{{Quark-gluon thermodynamics with the Z(N(c)) symmetry}},
  \href{https://doi.org/10.1088/0954-3899/39/8/085010}{\emph{J. Phys.}
  {\bfseries G39} (2012) 085010}.

\bibitem{Sakai:2012ika}
Y.~Sakai, H.~Kouno, T.~Sasaki and M.~Yahiro, \emph{{The quarkyonic phase and
  the Z$_{N_{c}}$ symmetry}},
  \href{https://doi.org/10.1016/j.physletb.2012.10.027}{\emph{Phys. Lett.}
  {\bfseries B718} (2012) 130--135},
  [\href{https://arxiv.org/abs/1204.0228}{{\ttfamily 1204.0228}}].

\bibitem{Kouno:2013zr}
H.~Kouno, T.~Makiyama, T.~Sasaki, Y.~Sakai and M.~Yahiro, \emph{{Confinement
  and $\mathbb{Z}_{3}$ symmetry in three-flavor QCD}},
  \href{https://doi.org/10.1088/0954-3899/40/9/095003}{\emph{J. Phys.}
  {\bfseries G40} (2013) 095003},
  [\href{https://arxiv.org/abs/1301.4013}{{\ttfamily 1301.4013}}].

\bibitem{Kouno:2013mma}
H.~Kouno, T.~Misumi, K.~Kashiwa, T.~Makiyama, T.~Sasaki and M.~Yahiro,
  \emph{{Differences and similarities between fundamental and adjoint matters
  in SU(N) gauge theories}},
  \href{https://doi.org/10.1103/PhysRevD.88.016002}{\emph{Phys. Rev.}
  {\bfseries D88} (2013) 016002},
  [\href{https://arxiv.org/abs/1304.3274}{{\ttfamily 1304.3274}}].

\bibitem{Kouno:2015sja}
H.~Kouno, K.~Kashiwa, J.~Takahashi, T.~Misumi and M.~Yahiro,
  \emph{{Understanding QCD at high density from a Z$_3$-symmetric QCD-like
  theory}}, \href{https://doi.org/10.1103/PhysRevD.93.056009}{\emph{Phys. Rev.}
  {\bfseries D93} (2016) 056009},
  [\href{https://arxiv.org/abs/1504.07585}{{\ttfamily 1504.07585}}].

\bibitem{Hirakida:2016rqd}
T.~Hirakida, H.~Kouno, J.~Takahashi and M.~Yahiro, \emph{{Interplay between
  sign problem and $Z_3$ symmetry in three-dimensional Potts models}},
  \href{https://doi.org/10.1103/PhysRevD.94.014011}{\emph{Phys. Rev.}
  {\bfseries D94} (2016) 014011},
  [\href{https://arxiv.org/abs/1604.02977}{{\ttfamily 1604.02977}}].

\bibitem{Larsen:2016fvs}
R.~Larsen and E.~Shuryak, \emph{{Instanton-dyon ensembles with quarks with
  modified boundary conditions}},
  \href{https://doi.org/10.1103/PhysRevD.94.094009}{\emph{Phys. Rev.}
  {\bfseries D94} (2016) 094009},
  [\href{https://arxiv.org/abs/1605.07474}{{\ttfamily 1605.07474}}].

\bibitem{Hirakida:2017bye}
T.~Hirakida, J.~Sugano, H.~Kouno, J.~Takahashi and M.~Yahiro, \emph{{Sign
  problem in $Z_3$-symmetric effective Polyakov-line model}},
  \href{https://arxiv.org/abs/1705.00665}{{\ttfamily 1705.00665}}.

\bibitem{Nye:2000eg}
T.~M.~W. Nye and M.~A. Singer, \emph{{An L**2 index theorem for Dirac operators
  on S**1 x R**3}}, {\emph{Submitted to: J. Funct. Anal.} (2000) },
  [\href{https://arxiv.org/abs/math/0009144}{{\ttfamily math/0009144}}].

\bibitem{Poppitz:2008hr}
E.~Poppitz and M.~Unsal, \emph{{Index theorem for topological excitations on
  R**3 x S**1 and Chern-Simons theory}},
  \href{https://doi.org/10.1088/1126-6708/2009/03/027}{\emph{JHEP} {\bfseries
  03} (2009) 027}, [\href{https://arxiv.org/abs/0812.2085}{{\ttfamily
  0812.2085}}].

\bibitem{Komargodski:2017smk}
Z.~Komargodski, T.~Sulejmanpasic and M.~{\"U}nsal, \emph{{Walls, Anomalies, and
  (De)Confinement in Quantum Anti-Ferromagnets}},
  \href{https://arxiv.org/abs/1706.05731}{{\ttfamily 1706.05731}}.

\bibitem{Vafa:1983tf}
C.~Vafa and E.~Witten, \emph{{Restrictions on Symmetry Breaking in Vector-Like
  Gauge Theories}},
  \href{https://doi.org/10.1016/0550-3213(84)90230-X}{\emph{Nucl. Phys.}
  {\bfseries B234} (1984) 173--188}.

\bibitem{Rosenzweig:1979ay}
C.~Rosenzweig, J.~Schechter and C.~G. Trahern, \emph{{Is the Effective
  Lagrangian for QCD a Sigma Model?}},
  \href{https://doi.org/10.1103/PhysRevD.21.3388}{\emph{Phys. Rev.} {\bfseries
  D21} (1980) 3388}.

\bibitem{DiVecchia:1980yfw}
P.~Di~Vecchia and G.~Veneziano, \emph{{Chiral Dynamics in the Large n Limit}},
  \href{https://doi.org/10.1016/0550-3213(80)90370-3}{\emph{Nucl. Phys.}
  {\bfseries B171} (1980) 253--272}.

\bibitem{Witten:1980sp}
E.~Witten, \emph{{Large N Chiral Dynamics}},
  \href{https://doi.org/10.1016/0003-4916(80)90325-5}{\emph{Annals Phys.}
  {\bfseries 128} (1980) 363}.

\bibitem{Witten:1978bc}
E.~Witten, \emph{{Instantons, the Quark Model, and the 1/n Expansion}},
  \href{https://doi.org/10.1016/0550-3213(79)90243-8}{\emph{Nucl. Phys.}
  {\bfseries B149} (1979) 285--320}.

\bibitem{Tanizaki:2017bam}
Y.~Tanizaki and Y.~Kikuchi, \emph{{Vacuum structure of bifundamental gauge
  theories at finite topological angles}},
  \href{https://doi.org/10.1007/JHEP06(2017)102}{\emph{JHEP} {\bfseries 06}
  (2017) 102}, [\href{https://arxiv.org/abs/1705.01949}{{\ttfamily
  1705.01949}}].

\bibitem{Kikuchi:2017pcp}
Y.~Kikuchi and Y.~Tanizaki, \emph{{Global inconsistency, 't~Hooft anomaly, and
  level crossing in quantum mechanics}},
  \href{https://arxiv.org/abs/1708.01962}{{\ttfamily 1708.01962}}.

\bibitem{Tanizaki:2017qhf}
Y.~Tanizaki, T.~Misumi and N.~Sakai, \emph{{Circle compactification and 't
  Hooft anomaly}},  \href{https://arxiv.org/abs/1710.08923}{{\ttfamily
  1710.08923}}.

\bibitem{Redlich:1983dv}
A.~N. Redlich, \emph{{Parity Violation and Gauge Noninvariance of the Effective
  Gauge Field Action in Three-Dimensions}},
  \href{https://doi.org/10.1103/PhysRevD.29.2366}{\emph{Phys. Rev.} {\bfseries
  D29} (1984) 2366--2374}.

\bibitem{Callias:1977kg}
C.~Callias, \emph{{Index Theorems on Open Spaces}},
  \href{https://doi.org/10.1007/BF01202525}{\emph{Commun. Math. Phys.}
  {\bfseries 62} (1978) 213--234}.

\bibitem{Csaki:2017cqm}
C.~Csaki, Y.~Shirman, J.~Terning and M.~Waterbury, \emph{{Twisted Sisters: KK
  Monopoles and their Zero Modes}},
  \href{https://arxiv.org/abs/1708.03330}{{\ttfamily 1708.03330}}.

\bibitem{DiPietro:2015taa}
L.~Di~Pietro, Z.~Komargodski, I.~Shamir and E.~Stamou, \emph{{Quantum
  Electrodynamics in d=3 from the ε Expansion}},
  \href{https://doi.org/10.1103/PhysRevLett.116.131601}{\emph{Phys. Rev. Lett.}
  {\bfseries 116} (2016) 131601},
  [\href{https://arxiv.org/abs/1508.06278}{{\ttfamily 1508.06278}}].

\bibitem{Giombi:2015haa}
S.~Giombi, I.~R. Klebanov and G.~Tarnopolsky, \emph{{Conformal QED$_d$,
  $F$-Theorem and the $\epsilon$ Expansion}},
  \href{https://doi.org/10.1088/1751-8113/49/13/135403}{\emph{J. Phys.}
  {\bfseries A49} (2016) 135403},
  [\href{https://arxiv.org/abs/1508.06354}{{\ttfamily 1508.06354}}].

\bibitem{Karthik:2015sgq}
N.~Karthik and R.~Narayanan, \emph{{No evidence for bilinear condensate in
  parity-invariant three-dimensional QED with massless fermions}},
  \href{https://doi.org/10.1103/PhysRevD.93.045020}{\emph{Phys. Rev.}
  {\bfseries D93} (2016) 045020},
  [\href{https://arxiv.org/abs/1512.02993}{{\ttfamily 1512.02993}}].

\bibitem{Karthik:2016ppr}
N.~Karthik and R.~Narayanan, \emph{{Scale-invariance of parity-invariant
  three-dimensional QED}},
  \href{https://doi.org/10.1103/PhysRevD.94.065026}{\emph{Phys. Rev.}
  {\bfseries D94} (2016) 065026},
  [\href{https://arxiv.org/abs/1606.04109}{{\ttfamily 1606.04109}}].

\bibitem{DiPietro:2017kcd}
L.~Di~Pietro and E.~Stamou, \emph{{Scaling dimensions in QED$_3$ from the
  $\epsilon$-expansion}},  \href{https://arxiv.org/abs/1708.03740}{{\ttfamily
  1708.03740}}.

\bibitem{Marston:1989zz}
J.~B. Marston and I.~Affleck, \emph{{Large-n limit of the Hubbard-Heisenberg
  model}}, \href{https://doi.org/10.1103/PhysRevB.39.11538}{\emph{Phys. Rev.}
  {\bfseries B39} (1989) 11538--11558}.

\bibitem{Rantner:2000wer}
W.~Rantner and X.-G. Wen, \emph{{Electron spectral function and algebraic spin
  liquid for the normal state of underdoped high $T_c$ superconductors}},
  \href{https://doi.org/10.1103/PhysRevLett.86.3871}{\emph{Phys. Rev. Lett.}
  {\bfseries 86} (2001) 3871},
  [\href{https://arxiv.org/abs/cond-mat/0010378}{{\ttfamily
  cond-mat/0010378}}].

\bibitem{Rantner:2002zz}
W.~Rantner and X.-G. Wen, \emph{{Spin correlations in the algebraic spin
  liquid: Implications for high-Tc superconductors}},
  \href{https://doi.org/10.1103/PhysRevB.66.144501}{\emph{Phys. Rev.}
  {\bfseries B66} (2002) 144501}.

\bibitem{Hermele:2005dkq}
M.~Hermele, T.~Senthil and M.~P.~A. Fisher, \emph{{Algebraic spin liquid as the
  mother of many competing orders}},
  \href{https://doi.org/10.1103/PhysRevB.72.104404}{\emph{Phys. Rev.}
  {\bfseries B72} (2005) 104404},
  [\href{https://arxiv.org/abs/cond-mat/0502215}{{\ttfamily
  cond-mat/0502215}}].

\bibitem{Hermele:2008afa}
M.~Hermele, Y.~Ran, P.~Lee and X.~G. Wen, \emph{{Properties of an algebraic
  spin liquid on the kagome lattice}},
  \href{https://doi.org/10.1103/PhysRevB.77.224413}{\emph{Phys. Rev.}
  {\bfseries B77} (2008) 224413}.

\bibitem{Ran:2007zea}
Y.~Ran, M.~Hermele, P.~A. Lee and X.~G. Wen, \emph{{Projected-Wave-Function
  Study of the Spin-1/2 Heisenberg Model on the Kagom{\'e} Lattice}},
  \href{https://doi.org/10.1103/PhysRevLett.98.117205}{\emph{Phys. Rev. Lett.}
  {\bfseries 98} (2007) 117205}.

\bibitem{TKMS_unpublished}
Y.~Kikuchi, T.~Misumi, N.~Sakai and Y.~Tanizaki, \emph{Anomaly matching for
  phase diagram of massless $\mathbb{Z}_n$-qcd}, {\emph{to appear} (2017) }.

\end{thebibliography}\endgroup
\bibliographystyle{JHEP}

\end{document}